\documentclass[longauth]{aa} 

\usepackage{graphicx}
\usepackage{multirow}
\usepackage{txfonts}
\usepackage{longtable}
\usepackage{comment}
\usepackage{subfig}
\usepackage{threeparttable}
\usepackage[colorlinks=true, citecolor=blue]{hyperref}
\usepackage{natbib}
\bibpunct{(}{)}{;}{a}{}{,} 
\RequirePackage{etex} 
\usepackage[english]{babel}
\usepackage{xcolor}
\usepackage{titletoc} 

\newcommand{\mstar}{$\rm M_\star$}
\newcommand{\mdot}{$\rm M_\odot$}

\defcitealias{Bethermin+15}{B15}
\defcitealias{Bing+23}{Bing23}
\defcitealias{Bethermin+26}{B26}

\begin{document} 

\title{The NIKA2 Cosmological Legacy Survey in COSMOS: Final 1.2\,mm and 2\,mm source catalogs and redshift distribution of dusty star-forming galaxies}

\author{C. R. Carvajal-Bohorquez\inst{\ref{LAM}}\corrauth{cristian.carvajal@lam.fr}
    \and G. Lagache\inst{\ref{LAM}}\corrauth{guilaine.lagache@lam.fr}
    \and A. Beelen\inst{\ref{LAM}}\email{alexandre.beelen@lam.fr}
     \and R.~Adam \inst{\ref{OCA}}\email{remi.adam@oca.eu}
     \and  P.~Ade \inst{\ref{Cardiff}}\email{AdePA@cardiff.ac.uk}
     \and  H.~Ajeddig \inst{\ref{CEA}}\email{hamza.ajeddig@cea.fr}
     \and  S.~Amarantidis \inst{\ref{IRAME}, \ref{Amaran1}, \ref{Amaran2}}\email{samarant@fc.ul.pt}
     \and  P.~Andr\'e \inst{\ref{CEA}}\email{philippe.andre@cea.fr}
     \and M. Aravena\inst{\ref{IEA},\ref{MINGAL}}\email{manuel.aravenaa@mail.udp.cl}
     \and  H.~Aussel \inst{\ref{CEA}}\email{herve.aussel@cnrs.fr}
     \and  A.~Beno\^it \inst{\ref{Neel}}\email{alain.benoit@neel.cnrs.fr}
    \and S.~Berta \inst{\ref{IRAMF}}\email{berta@iram.fr}
     \and M.~B\'ethermin \inst{\ref{UniStra}}\email{matthieu.bethermin@astro.unistra.fr}
     \and L.-J.~Bing\inst{\ref{Sussex}}\email{binglj94@gmail.com}
     \and  A.~Bongiovanni \inst{\ref{IRAME}}\email{bongio@iram.es}
     \and  J.~Bounmy \inst{\ref{LPSC}}\email{bounmy@lpsc.in2p3.fr}
     \and  O.~Bourrion \inst{\ref{LPSC}}\email{bourrion@lpsc.in2p3.fr}
     \and  M.~Calvo \inst{\ref{Neel}}\email{martino.calvo@neel.cnrs.fr}
     \and Caitlin~M.Casey\inst{\ref{Santa_barbara},\ref{dawn}}\email{cmcasey@ucsb.edu}
     \and  A.~Catalano \inst{\ref{LPSC}}\email{catalano@lpsc.in2p3.fr}
     \and  D.~Ch\'erouvrier \inst{\ref{LPSC}}\email{cherouvrier@lpsc.in2p3.fr}
     \and  U.~Chowdhury \inst{\ref{Neel}}\email{usasi.chowdhury@neel.cnrs.fr}   
     \and  M.~De~Petris \inst{\ref{Roma}}\email{marco.depetris@roma1.infn.it}
     \and  F.-X.~D\'esert \inst{\ref{IPAG}}\email{francois-xavier.desert@univ-grenoble-alpes.fr}
     \and  S.~Doyle \inst{\ref{Cardiff}}\email{DoyleSM@cardiff.ac.uk}
     \and  E.~F.~C.~Driessen \inst{\ref{IRAMF}}\email{driessen@iram.fr}
     \and  G.~Ejlali \inst{\ref{Teheran}}\email{gejlali@ipm.ir}
     \and A.L.~Faisst\inst{\ref{caltech}}\email{afaisst@ipac.caltech.edu}
     \and  A.~Ferragamo \inst{\ref{Roma}}\email{antonio.ferragamo@uniroma1.it}
     \and  A.~Gomez \inst{\ref{CAB}}\email{gomezgf@cab.inta-csic.es}
     \and  J.~Goupy \inst{\ref{Neel}}\email{johannes.goupy@neel.cnrs.fr}
     \and  C.~Hanser \inst{\ref{CPPM}}\email{hanser@cppm.in2p3.fr}
     \and Shuowen Jin\inst{\ref{dawn}, \ref{dtu}}\email{shuji@dtu.dk}
     \and Jeyhan~S.~Kartaltepe\inst{\ref{Rochester}}\email{jsksps@rit.edu}
     \and  S.~Katsioli \inst{\ref{AthenObs}, \ref{AthenUniv}}\email{s.katsioli@noa.gr}
     \and  F.~K\'eruzor\'e \inst{\ref{Argonne}}\email{fkeruzore@anl.gov}
     \and  C.~Kramer \inst{\ref{IRAMF}}\email{kramer@iram.fr}
     \and  B.~Ladjelate \inst{\ref{IRAME}}\email{ladjelate@iram.es}
     \and  S.~Leclercq \inst{\ref{IRAMF}}\email{leclercq@iram.fr}
     \and  J.-F.~Lestrade \inst{\ref{LUX}}\email{Jean-Francois.Lestrade@obspm.fr}
     \and  J.~F.~Mac\'ias-P\'erez \inst{\ref{LPSC}}\email{juan.macias-perez@lpsc.in2p3.fr}
     \and  S.~C.~Madden \inst{\ref{CEA}}\email{suzanne.madden@cea.fr}
     \and Felix~Martinez~III\inst{\ref{Rochester}}\email{felixmartz123@gmail.com}
     \and  A.~Maury \inst{\ref{Barcelona1}, \ref{Barcelona2}, \ref{CEA}}\email{anaelle.maury.astro@gmail.com}
     \and  F.~Mayet \inst{\ref{LPSC}}\email{frederic.mayet@lpsc.in2p3.fr}
     \and  A.~Monfardini \inst{\ref{Neel}}\email{alessandro.monfardini@neel.cnrs.fr}
     \and  A.~Moyer-Anin \inst{\ref{LPSC}}\email{alice.moyer@lpsc.in2p3.fr}
     \and  M.~Mu\~noz-Echeverr\'ia \inst{\ref{IRAP}}\email{mirenmunoze@gmail.com}
     \and  I.~Myserlis \inst{\ref{IRAME}}\email{imyserlis@iram.es}
     \and R.~Neri\inst{\ref{IRAMF}}\email{neri@iram.fr}
     \and  A.~Paliwal \inst{\ref{Roma2}}\email{Aishwarya.Paliwal@roma2.infn.it}
     \and  L.~Perotto \inst{\ref{LPSC}}\email{laurence.perotto@lpsc.in2p3.fr}
     \and  G.~Pisano \inst{\ref{Roma}}\email{giampaolo.pisano@astro.cf.ac.uk}
     \and  N.~Ponthieu \inst{\ref{IPAG}}\email{nicolas.ponthieu@univ-grenoble-alpes.fr}
     \and  V.~Rev\'eret \inst{\ref{CEA}}\email{vincent.reveret@cea.fr}
     \and  A.~J.~Rigby \inst{\ref{Leeds}}\email{ajrigby24@gmail.com}
     \and  A.~Ritacco \inst{\ref{LPSC}}\email{Alessia.Ritacco@lpsc.in2p3.fr}
     \and  H.~Roussel \inst{\ref{IAP}}\email{roussel@iap.fr}
     \and  F.~Ruppin \inst{\ref{IP2I}}\email{f.ruppin@ip2i.in2p3.fr}
     \and  M.~S\'anchez-Portal \inst{\ref{IRAME}}\email{msanchez@iram.es}
     \and  S.~Savorgnano \inst{\ref{LPSC}}\email{sofia.savorgnano@lpsc.in2p3.fr}
     \and  K.~Schuster \inst{\ref{IRAMF}}\email{schuster@iram.fr}
     \and  A.~Sievers \inst{\ref{IRAME}}\email{sievers@iram.es}
     \and S.~Toft\inst{\ref{dark}}\email{sune@nbi.ku.dk}
     \and  C.~Tucker \inst{\ref{Cardiff}}\email{carole.tucker@astro.cf.ac.uk}
     \and Jorge A. Zavala\inst{\ref{UMA}}\email{jzavala@umass.edu} 
     \and  R.~Zylka \inst{\ref{IRAMF}}\email{zylka@iram.fr}
     }

   \institute{
    Aix Marseille Univ, CNRS, CNES, LAM (Laboratoire d'Astrophysique de Marseille), Marseille, France
    \label{LAM}
         \and
    Universit\'e C\^ote d'Azur, Observatoire de la C\^ote d'Azur, CNRS, Laboratoire Lagrange, France 
     \label{OCA}
    \and
     School of Physics and Astronomy, Cardiff University, Queen’s Buildings, The Parade, Cardiff, CF24 3AA, UK 
     \label{Cardiff}
     \and
     Université Paris Cité, Université Paris-Saclay, CEA, CNRS, AIM, F-91191 Gif-sur-Yvette, France
     \label{CEA}
    \and
     Institut de Radioastronomie Millim\'etrique (IRAM), Avenida Divina Pastora 7, Local 20, E-18012, Granada, Spain
     \label{IRAME}
     \and
     Instituto de Astrofísica e Ciências do Espaço, Universidade de Lisboa, OAL, Tapada da Ajuda, PT1349-018 Lisbon, Portugal
    \label{Amaran1}
     \and
    Departamento de Física, Faculdade de Ciências, Universidade de Lisboa, Edifício C8, Campo Grande, PT1749-016 Lisbon, Portugal
     \label{Amaran2}
     \and
     Instituto de Estudios Astrof\'{\i}cos, Facultad de Ingenier\'{\i}a y Ciencias, Universidad Diego Portales, Av. Ej\'ercito 441, Santiago, Chile
    \label{IEA}
    \and
    Millennium Nucleus for Galaxies (MINGAL)
    \label{MINGAL} 
     \and
     Institut N\'eel, CNRS, Universit\'e Grenoble Alpes, France
     \label{Neel}
     \and
     Institut de Radioastronomie Millim\'etrique (IRAM), 300 rue de la Piscine, 38400 Saint-Martin-d'H{\`e}res, France
     \label{IRAMF}
    \and
    Universit\'e de Strasbourg, CNRS, Observatoire astronomique de Strasbourg, UMR 7550, 67000 Strasbourg, France
     \label{UniStra}
     \and
    Astronomy Centre, Department of Physics and Astronomy, University of Sussex, Brighton BN1 9QH
     \label{Sussex}
    \and
      Univ. Grenoble Alpes, CNRS, Grenoble INP, LPSC-IN2P3, 53, avenue des Martyrs, 38000 Grenoble, France
     \label{LPSC}
     \and
    Department of Physics, University of California, Santa Barbara, Santa Barbara, CA 93106, USA
    \label{Santa_barbara}
    \and 
    Cosmic Dawn Center (DAWN), Denmark
    \label{dawn}
     \and 
     Dipartimento di Fisica, Sapienza Universit\`a di Roma, Piazzale Aldo Moro 5, I-00185 Roma, Italy
     \label{Roma}
     \and
     Univ. Grenoble Alpes, CNRS, IPAG, 38000 Grenoble, France 
     \label{IPAG}
     \and
     Institute for Research in Fundamental Sciences (IPM), School of Astronomy, Tehran, Iran
     \label{Teheran}
     \and
     IPAC, California Institute of Technology, 1200 E. California Blvd., Pasadena, CA 91125, USA
        \label{caltech}  
     \and
     Centro de Astrobiolog\'ia (CSIC-INTA), Torrej\'on de Ardoz, 28850 Madrid, Spain
     \label{CAB}
     \and
     Aix Marseille Univ, CNRS/IN2P3, CPPM, Marseille, France
     \label{CPPM}
    \and
     DTU Space, Technical University of Denmark, Elektrovej 327, DK-2800 Kgs. Lyngby, Denmark 
     \label{dtu}
    \and
     Laboratory for Multiwavelength Astrophysics, School of Physics and Astronomy, Rochester Institute of Technology, 84 Lomb Memorial Drive, Rochester, NY 14623, USA
     \label{Rochester}
     \and
     National Observatory of Athens, Institute for Astronomy, Astrophysics, Space Applications and Remote Sensing, Ioannou Metaxa
     and Vasileos Pavlou GR-15236, Athens, Greece
     \label{AthenObs}
     \and
     Department of Astrophysics, Astronomy \& Mechanics, Faculty of Physics, University of Athens, Panepistimiopolis, GR-15784
     Zografos, Athens, Greece
     \label{AthenUniv}
     \and
     High Energy Physics Division, Argonne National Laboratory, 9700 South Cass Avenue, Lemont, IL 60439, USA
     \label{Argonne}
     \and  
     LUX, Observatoire de Paris, PSL Research University, CNRS, Sorbonne Universit\'e, UPMC, 75014 Paris, France  
     \label{LUX}
     \and
     Institute of Space Sciences (ICE), CSIC, Campus UAB, Carrer de Can Magrans s/n, E-08193, Barcelona, Spain
     \label{Barcelona1}
     \and
     ICREA, Pg. Lluís Companys 23, Barcelona, Spain
     \label{Barcelona2}
     \and
     IRAP, CNRS, Université de Toulouse, CNES, UT3-UPS, (Toulouse), France
     \label{IRAP}
    \and
     Dipartimento di Fisica, Universit\`a di Roma ‘Tor Vergata’, Via della Ricerca Scientifica 1, I-00133 Roma, Italy        
     \label{Roma2}
     \and
     School of Physics and Astronomy, University of Leeds, Leeds LS2 9JT, UK
     \label{Leeds}
     \and    
     Institut d'Astrophysique de Paris, CNRS (UMR7095), 98 bis boulevard Arago, 75014 Paris, France
     \label{IAP}
     \and
     University of Lyon, UCB Lyon 1, CNRS/IN2P3, IP2I, 69622 Villeurbanne, France
     \label{IP2I}
     \and
     Dark Cosmology Centre, Niels Bohr Institute, University of Copenhagen, Juliane Maries Vej 32, København Ø, 2100, Denmark
     \label{dark}
     \and
     University of Massachusetts Amherst, 710 North Pleasant Street, Amherst, MA 01003-9305, USA
     \label{UMA}
}
   \date{Received 30 April 2026; accepted 06 October 2026}
   
 \titlerunning{N2CLS COSMOS}
 \authorrunning{C. R. Carvajal-Bohorquez et al.}

 \abstract
{Dusty star-forming galaxies (DSFGs) play a major role in the overall star formation in the Universe at high redshift. While deep near-infrared surveys, particularly with JWST, can now identify many heavily obscured galaxies, (sub)millimeter observations are essential for measuring their dust emission and quantifying their dust content and obscured star formation.}{We present the final 1.2\,mm and 2\,mm source catalogs and the redshift distribution of the millimeter-selected population from the NIKA2 Cosmological Legacy Survey (N2CLS) in the COSMOS field. Our aim was to provide a comprehensive dataset for studying the physical properties and evolution of high-redshift DSFGs and to compare the observed redshift distribution with theoretical predictions.}{We analyzed deep NIKA2 observations at 1.2\,mm and 2\,mm, covering $\sim$1070\,arcmin$^2$ with median noise levels of 315\,$\mathrm \mu$Jy and 91\,$\mathrm \mu$Jy, respectively. Sources were extracted with a signal-to-noise ratio (S/N) threshold of 3.9, ensuring $\geq$80\% purity. We  identified multiwavelength counterparts using high-resolution interferometric (sub)millimeter data (NOEMA, ALMA) and radio observations (VLA, MeerKAT). Redshifts were compiled from spectroscopic and photometric catalogs (e.g., COSMOSWeb). We compared the redshift distribution to predictions from the SIDES simulations and other galaxy evolution models (e.g., SHARK and FLAMINGO).}{The N2CLS master catalog includes 323 sources detected at $\geq$80\% purity, with 104 sources detected in both bands, 197 only at 1.2\,mm, and 22 only at 2\,mm. Multiwavelength identifications were secured for $\sim$89\% of the sample. The redshift distribution of 1.2\,mm high-purity ($\geq95\%$) sources peaks at $2.8 \pm 0.1$, consistent with the epoch of peak cosmic star formation. 
In the total sample, we lack redshifts for $\sim2\%$ of the identified galaxies plus 34 sources for which no accurate positional proxy is available, preventing the identification of a multiwavelength counterpart.
We identified 66 galaxies at $z > 4$ (19 with spectroscopic redshifts). The observed redshift distribution agrees well with the SIDES simulations, while four other galaxy evolution models are statistically inconsistent with the data.}{Using NIKA2 at the IRAM 30\,m telescope, we conducted N2CLS, the largest contiguous deep millimeter survey to date at this depth with homogeneous coverage. This homogeneous coverage is important, as 25\% of N2CLS sources lack a SCUBA2 850\,$\rm \mu m$ counterpart, likely because the strongly nonuniform noise distribution of the SCUBA2 map results in lower sensitivity in parts of the field.
The released data products, including maps, catalogs, and multiwavelength identifications, provide a legacy dataset for studying dust-obscured galaxy evolution. The N2CLS bridges wide-area single-dish observations and high-resolution interferometric studies, enabling future investigations into the physical properties and cosmic evolution of DSFGs.}


\maketitle 
\nolinenumbers 

\section{Introduction}
Dusty star-forming galaxies (DSFGs) play a central role in our understanding of galaxy formation and evolution, as they host a substantial fraction of the cosmic star formation activity at high redshift while being obscured at ultraviolet and optical wavelengths  \cite[e.g.,][]{Blain2002, Lagache2005, Casey2014}. Over the past two decades, wide-area surveys at (sub)millimeter wavelengths have revealed a population of luminous dust-obscured galaxies, often referred to as submillimeter galaxies, that dominate the bright end of the far-infrared luminosity function at redshifts z$\sim$1–5 \cite[e.g.,][]{Chapman+05, Simpson+14, Koprowski2017, Dudzeviciute2020, Zavala2021, Traina2024}. These systems are now recognized as key contributors to the cosmic star formation rate density (SFRD) during the peak epoch of galaxy assembly. At higher redshifts, recent Atacama Large Millimeter/submillimeter Array (ALMA) surveys (e.g., ALPINE and REBELS) have shown that dust-obscured (far-IR, millimeter) and unobscured (UV rest-frame) star formation could have a similar contribution to the cosmic budget even at z$\sim$6 \citep[e.g.,][]{Gruppioni2020, Fudamoto2020, Khusanova2021, Barrufet2023, fujimoto2024, Traina2024}. 
One of the main contributors to the obscured SFRD at such high redshifts is (sub)mm-bright ultra massive and heavily obscured (so-called optically dark) galaxies \cite[e.g.,][]{alcade2019, wang2019, Barrufet2023, xiao2023, Gottumukkala2024, xiao2024,lagache2026}. In a recent analysis, \cite{Manning2025} showed that such galaxies potentially contribute up to 60\% at the very high mass end (log$_{10}$(\mstar/\mdot)$>11.5$) of the $z>4$ stellar mass function. 

Early discoveries of submillimeter galaxies were enabled by single-dish instruments such as SCUBA and SCUBA2 on the \textit{James Clerk Maxwell} Telescope \citep[JCMT;][]{Holland1999, Holland2013}, LABOCA on the Atacama Pathfinder EXperiment \citep[APEX;][]{Weiss+09}, and AzTEC on several facilities \citep{Scott2008, Aretxaga2011}. While these surveys established the basic demographics of DSFGs, their coarse angular resolution introduced significant challenges for flux deboosting, source identification, and counterpart association \citep[e.g.,][]{Hodge2013}. This motivated extensive interferometric follow-up campaigns with facilities such as ALMA and the NOrthern Extended Millimeter Array (NOEMA), which revealed that a substantial fraction of the millimeter-bright single-dish sources break up into multiple components and that the population spans a wide range of redshifts and intrinsic luminosities \citep[e.g.,][]{Karim2013, Simpson2015, Stach+19, Berta+25}.

The advent of large-area multiwavelength extragalactic fields, most notably the Cosmic Evolution Survey \citep[COSMOS; RA=10:00:28.6, Dec=+02:12:21.0,][]{Scoville+07}, has transformed the study of DSFGs by enabling robust source identification and redshift estimation for statistically significant samples. COSMOS provides deep and homogeneous coverage from X-ray to radio wavelengths, making it a reference field for studies of galaxy evolution and large-scale structure across cosmic time. Numerous submillimeter and millimeter surveys have targeted COSMOS, including SCUBA2 \citep{Casey2013, Geach2017, Simpson+19, Gao2024} on JCMT, AzTEC \citep{Aretxaga2011} on the Atacama Submillimeter Telescope Experiment (ASTE), and ALMA blank-field and follow-up programs (e.g., \citealt{Simpson2015}, \citealt{Jin+19}, \citealt{Long2026}, \citealt{Zavala2026}, \citealt{Faisst2026}).

In this context, the NIKA2 camera on the IRAM (Institut de Radioastronomie Millimétrique) 30-m telescope \citep{Adam2018, Perotto2020} provides a powerful complement to existing surveys. Operating simultaneously at 1.2 and 2\,mm, NIKA2 combines a relatively high angular resolution ($\sim$11$\arcsec$ at 1.2\,mm) with the ability to map wide areas efficiently at millimeter wavelengths. Selection at millimeter wavelengths probes the dust emission further near the Rayleigh–Jeans tail of the infrared spectral energy distribution (SED).
While the k-correction remains negative from submillimeter to millimeter wavelengths, it is strongest around 850\,$\mu$m and becomes progressively weaker at longer wavelengths \citep[e.g.,][]{Blain2002, Casey2014, Bethermin+15}. As a result, millimeter surveys remain sensitive to high-redshift DSFGs but increasingly disfavor lower-redshift systems \citep{Bethermin+26}.
Millimeter surveys therefore provide a complementary view of DSFG populations selected under different SEDs and redshift sensitivities than traditional submillimeter surveys.

The NIKA2 Cosmological Legacy Survey (N2CLS) is a large program designed to exploit the capabilities of the NIKA2 camera by mapping two well-studied extragalactic fields  (GOODS-N and COSMOS) at 1.2 and 2\,mm. The survey and the first measurements of millimeter galaxy number counts\footnote{Based on a preliminary dataset corresponding to about half of the COSMOS observations.} are presented in \cite{Bing+23}, while subsequent companion papers have analyzed different aspects of the dataset, including the data products (e.g., multiwavelength N2CLS catalog) and the evolution of the cosmic dust density in GOODS-N \citep{Berta+25}, the statistical properties of the 2 mm-selected galaxy population in both GOODS-N and COSMOS \citep{Bethermin+26}, the analysis of the z$\sim$5.2 overdensity of DSFGs in GOODS-N \citep{lagache2026}, and the confusion noise limits of the IRAM 30 m telescope derived from the GOODS-N deep survey observations \citep{Ponthieu2026}.

This paper presents the final data reduction of the NIKA2 observations on the COSMOS field obtained as part of N2CLS together with the associated source catalogs, the identification products, and the resulting redshift distribution. The COSMOS field is a primary target due to its exceptional ancillary data, and its value has been further enhanced by recent public releases, including deep JWST imaging and associated catalogs \citep{shuntov+25}.
The primary goal of our work is to provide a detailed and homogeneous description of the dataset and of the analysis steps required to construct a robust sample of millimeter-selected galaxies. 
The present paper is intentionally focused on data presentation, validation, and catalog release. A companion paper (Berta et al., in prep.) will exploit this dataset to perform a detailed analysis of the SEDs of the sources and to derive statistical constraints on their physical properties.

The paper is organized as follows. Section~\ref{sec: data_red} describes the NIKA2 observations of the COSMOS field, the data reduction process, the source extraction methodology, purity estimates, and flux density corrections. Section~\ref{sec: N2CLS_catalog} presents the final N2CLS source catalog. Section~\ref{sec: multiwave_ident} details the identification of N2CLS sources, their counterparts at other wavelengths, and the construction of
their SEDs. In Sect.~\ref{sec:z_distrib}, we present the redshift distribution of the identified sources and compare it with predictions from existing phenomenological and semi-empirical models of DSFG populations \citep[e.g.,][]{Bethermin+17, Lagos2018, Popping+20, Zavala2021}.
In Sect.~\ref{sec:discussion}, we discuss the unidentified N2CLS sources and the 
SCUBA2-dark sources. Finally, in Sect.~\ref{sec:conclusion} we summarize the main results and outline prospects for future analyses.

\section{The N2CLS in the COSMOS field \label{sec: data_red}}

\subsection{Observations}
The COSMOS field was observed with NIKA2 from February 2018 to January 2023 (project ID 192-16). A total of 878 observations were obtained, excluding aborted or canceled scans, corresponding to an overall integration time of 192.81 hours. We conducted two sets of orthogonal on-the-fly scans, measuring 27.0\arcmin$\times$34.7\arcmin\ and 35.0\arcmin$\times$28.0\arcmin, respectively, centered at RA=10:00:28.81 and DEC=02:17:30.44.
The two groups of scans were performed at a scanning speed of 60\,arcsec\,s$^{-1}$, with position angles of 0$^\circ$ and +90$^\circ$ in the RA-Dec coordinate system. The observations were obtained with a median line-of-sight opacity of 0.24 at 255\,GHz, and a median absolute deviation of 0.08. A total of 28, 23, and 29 corrupted scans were excluded from the analysis for arrays 1 (1.2\,mm), 2 (2\,mm), and 3 (1.2\,mm), respectively.

\citealt{Bing+23} \citepalias[hereafter][]{Bing+23} introduced the N2CLS survey and presented 1.2 and 2\,mm number-count measurements based on tiered N2CLS observations covering 160\,arcmin$^{2}$ in GOODS-N and 1010\,arcmin$^{2}$ in COSMOS. For COSMOS, only data obtained prior to May 2021 were used. The final dataset presented here covers exactly the same area but has an integration time 2.3 times larger, resulting in a depth that is 1.5 times deeper. 

\begin{figure*}[!t]
\centering
\includegraphics[width=18cm]{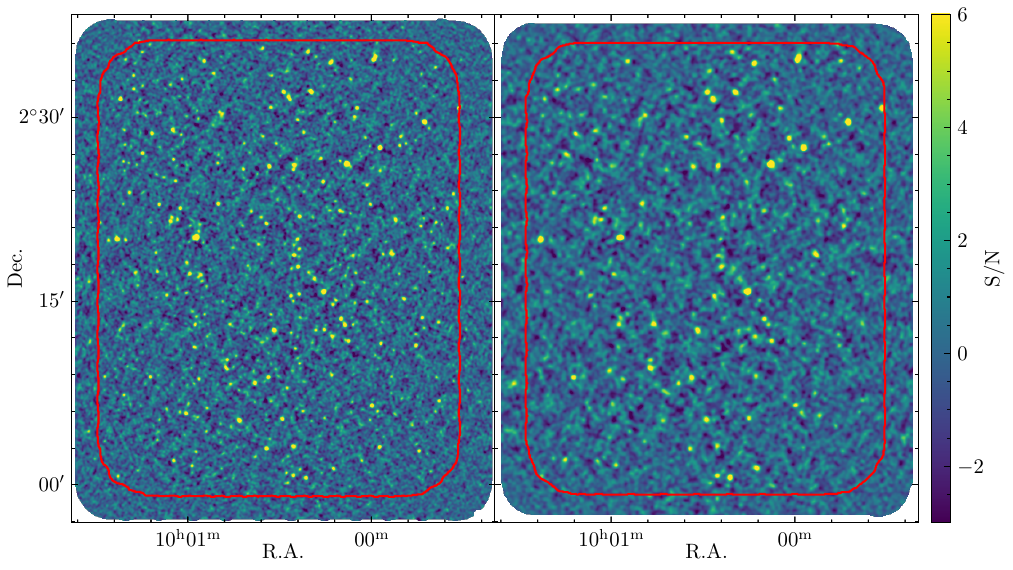}
\caption{N2CLS COSMOS maps of the signal-to-noise ratio at 1.2\,mm (\textit{left}) and 2\,mm (\textit{right}). The S/N maps were match filtered with the beam ($\rm FWHM=11.6\arcsec$ at 1.2\,mm and $\rm FWHM=18\arcsec$ at 2\,mm). 
The region enclosed in the red contour represents $1070$\,arcmin$^2$ and corresponds to the area where the sources were extracted.}
\label{fig: maps_1.2_2_mm}
\end{figure*}

\begin{table}
\centering
\caption{Main parameters used in the PIIC data reduction pipeline.}
\label{tab:piic}
\begin{tabular}{lll}
\hline
Parameter & Value & Description \\
\hline
\textit{weakSou}, \textit{deepField} & yes & Optimized configuration for \\
&  & faint-source detection \\
\textit{blOrderOrig} & 17 & Polynomial order for baseline \\
& & subtraction \\
\textit{nBest} & 12 & Number of best correlated \\ 
& & KIDs used for sky removal \\
\textit{nIterSource} & 10 & Iterative refinement of sky \\
&&  emission removal \\
\textit{snrLevel} & 4 & Threshold used to build the \\
& & source mask \\
\textit{souSign} & 1 & Sources constrained to be \\
& & positive in the mask \\
\hline
\end{tabular}
\end{table}

\subsection{Data reduction and maps} \label{subsec: data_redu}

The N2CLS COSMOS data were reduced using the \textit{jul23a}\footnote{\url{https://www.iram.fr/~gildas/dist/archive/piic/piic-exe-jul23a.tar.xz}.} version of the PIIC (Pointing and Imaging In Continuum) data reduction pipeline developed and supported by IRAM \citep{Zylka2013,Berta20192025}. We adapted the PIIC template script to run on a SLURM cluster using a map-reduce architecture with up to 99 cores. This allowed for a reduction speed-up allowing exploration of reduction parameters. Following \citetalias{Bing+23}, we adopted the pipeline configuration summarized in Table~\ref{tab:piic}.
We  used the PIIC iterative procedure, where a source map model is removed from the timeline, with 10 iterations to improve on the estimation and removal of the sky emission. The source map is constructed by combining all the scans and applying a $\ge$4$\sigma$ thresholding.

We have in total 850+849 and 855 scans that were projected on the sky at 1.2 and 2\,mm, respectively. Figure~\ref{fig: maps_1.2_2_mm} presents the signal-to-noise ratio (S/N) maps. The maps were match filtered using the beam corresponding to each band, i.e., full width at half maximum $\rm FWHM = 11.6\arcsec$ for 1.2\,mm and $\rm FWHM = 18\arcsec$ for 2\,mm. The pixel sizes of the maps are $3\arcsec$ and $4\arcsec$ at 1.2 and 2\,mm, respectively. Due to the elevated noise levels near the map edges (Fig.\,\ref{fig:noise_maps}), source extraction is restricted to the region enclosed by the red contour, which defines the area where the instrumental noise ($\sigma_{inst}$) is less than 1.8 
times the minimum noise value at the field center. This threshold represents a balance between ensuring noise homogeneity and maximizing the usable survey area. The resulting extraction region spans 1070\,arcmin$^2$. We thus extend the survey area by 60\,arcmin$^2$ with a lower $\sigma_{\rm inst}$ compared to \citetalias{Bing+23}.  
The median noise levels are 315\,$\mathrm{\mu Jy}$/beam
at 1.2\,mm and 91\,$\mathrm{\mu Jy}$/beam at 2\,mm.  
The noise maps, shown in Fig.\,\ref{fig:noise_maps}, are highly homogeneous within the source extraction area, with standard deviations of 60 and 20\,$\mathrm{\mu Jy}$/beam at 1.2 and 2\,mm, respectively.

\begin{figure}
\centering
\includegraphics[width=1\linewidth]{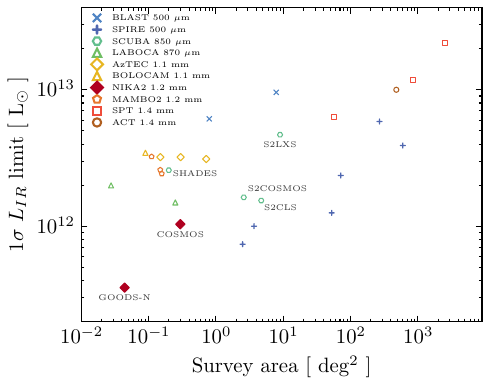}
\caption{Sensitivity–area comparison for representative single-dish (sub)millimeter extragalactic single-dish surveys. The plot shows the 1$\sigma$ infrared luminosity limit (L$_{\rm IR}$ in L$_{\odot}$) as a function of survey area (deg$^2$) for a galaxy at z=2.8 using the SED main-sequence template from SIDES \citep{Bethermin+17}. Different symbols indicate surveys conducted at various observing wavelengths, including 500\,$\rm \mu$m BLAST \citep{Dye+09} and SPIRE \citep{Oliver+12}, 850–870\,$\rm \mu$m SCUBA/SCUBA2 \citep{Mortier+05,Geach2017,Simpson+19,Garratt+23} and LABOCA \citep{Weiss+09}, 1.1\,mm AzTEC \citep{Scott2008,Aretxaga2011} and BOLOCAM \citep{Laurent+05}, 1.2\,mm NIKA2 and MAMBO2 \citep{Bertoldi+07,Lindner+11}, and 1.4\,mm SPT \citep{Mocanu+13} and ACT \citep{Gralla+20}. Several well-known surveys are labeled (e.g., SHADES, S2COSMOS, S2CLS, S2LXS). The figure illustrates the typical trade-off between survey depth and area.}
\label{fig:Lir_area}
\end{figure}

Figure\,\ref{fig:Lir_area} shows the $1\sigma$ infrared luminosity limit $L_{\mathrm{IR}}$ as a function of survey area for representative single-dish (sub)millimeter surveys conducted between 500\,$\mu$m and 1.4\,mm. The red diamonds indicate the location of the N2CLS observations in the GOODS-N and COSMOS fields. Assuming $z=2.8$ (median redshift of N2CLS 1.2\,mm sources; see Sect.~\ref{sec:n2cls_galaxies}), the NIKA2 COSMOS survey probes $\sim0.3\,\mathrm{deg}^2$ with $1\sigma$ luminosity limits of $10^{12}\,L_\odot$, occupying an intermediate regime between ultra-deep surveys and much wider but shallower single-dish surveys. As such, the NIKA2 COSMOS observations provide a complementary view of the DSFG population by simultaneously sampling a statistically meaningful area while retaining sufficient depth to detect galaxies with $L_{\mathrm{IR}}$ characteristic of luminous infrared galaxies at high redshift.\\

Following \citetalias{Bing+23}, we generated 100 half-difference maps from the full set of original scans by randomly dividing the scans into two subsets with opposite signs and co-adding them. These maps provide noise realizations that closely reproduce the instrumental noise properties of the observations. The same procedure is applied when injecting simulated input maps into the KID timelines to produce realistic mock observations.

\subsection{Source extraction}\label{subsec: source_extraction}
We followed the approach described in Sect. 2.3 of \citetalias{Bing+23}, with a few modifications, for source detection and flux measurement. We refer the reader to that paper for full details. Here we summarize only the main steps and highlight the changes introduced in this work.

First, we rescaled the weight map to ensure that the resulting S/N distribution has a standard deviation of unity. For that purpose, we perform a Gaussian fitting on the pixel S/N histogram values between -3 and 1.5 to avoid the contamination by the sources.  Furthermore, since the absolute level is undefined in the PIIC maps, we applied a correction to enforce a zero mean level. To that end, we implemented a routine to measure the mean level of the signal distribution, estimating the peak by iteratively fitting a Gaussian to the central peak of the distribution. We correct the zero-level on the signal maps using the resulting value. Finally, as source detection is performed on the match-filtered PIIC maps, the match-filtered weight map was rescaled to obtain a normalized match-filtered S/N map.

We detected sources on the match-filtered maps using the \texttt{find\_peak} algorithm from photutils \citep{Bradley+22}, which is incorporated into nikamap \citep{beelen_alexandre_2023_7520530}, setting an S/N threshold according to the purity (see Sect.~\ref{sec: purity}). To measure the flux, we performed point spread function (PSF) fitting photometry with nikamap using the \texttt{BasicPSFPhotometry} module from photutils, in the original PIIC maps at the positions given by \texttt{find\_peak}. 

\subsection{Catalog characterization \label{sect:cat_char}}
To build a robust catalog from far-IR and millimeter blind surveys, it is crucial to quantify the effects that impact source detection and flux measurements, and to account for systematic effects introduced during data reduction. Previous works have employed methods with statistical or empirical approaches (e.g., \citealt{Coppin+05} and \citealt{Zavala2021}, respectively) to estimate deviations in flux measurement, false detection rate, or completeness. However, these approaches focus mainly on instrumental noise and neglect astrophysical effects such as clustering and random alignments of sources inside the beam, which have been shown to significantly affect source fluxes in single-dish observations at long wavelengths \citep{Bethermin+17}.

We used the simulated sky maps produced in \citetalias{Bing+23} based on the Simulated Infrared Dusty Extragalactic Sky simulation \citep[SIDES,][]{Bethermin+17, Bethermin+22, Gkogkou+23} as our noiseless sky model. 
The 1.2 and 2\,mm fluxes are  derived using the NIKA2 bandpass, and maps are produced based on the fluxes and positions of all galaxies. The maps are then convolved with the two-dimensional NIKA2 Gaussian beam.
This allowed us to account for the effects of galaxy clustering and confusion in our catalog.
The 117 deg$^2$ of SIDES are divided into 117 tiles of 1 deg$^2$ to produce the 117 sky maps for each wavelength.

To account for instrumental noise,  pipeline artifacts, and astrophysical effects, we inject the simulated sky maps into the timelines using PIIC. 
To this end, we used a beam-convolved, noiseless sky model. Given as an input to PIIC, this model is projected back into the individual kinetic inductance detector (KID) timelines for all the scans, using a half-difference scheme described in Sect.\,\ref{subsec: data_redu}.
In this way, we generate simulated NIKA2 observations from the simulated sky maps with the same characteristics, such as noise, depth, and similar signal filtering.
For further details on this methodology, we refer the reader to \citetalias{Bing+23}, their Sect.~3.1.

Applying this approach to the 117 simulated sky maps, we generated two different types of catalogs. The ``simulated catalogs'' were extracted from the simulated observations (N2CLS-like maps), following the procedure explained in Sect.~\ref{subsec: source_extraction}; therefore, they are equivalent to catalogs derived from real observations. We also built ``blob catalogs,'' which correspond to sources extracted from the noiseless beam convolved sky maps. As these maps do not include noise, we introduced specific changes to the source detection process. To estimate the zero mean level, we fit a Log-normal distribution to the signal distribution to measure the peak position, which is used to set the zero level. Source detection is performed directly on the noiseless signal maps by applying flux thresholds. The thresholds are set to the median noise level corresponding to
315\,$\mathrm \mu$Jy and 91$\mathrm \mu$Jy at 1.2 and 2\,mm, respectively. 
We measured the fluxes using PSF-fitting photometry at the detected source positions. By construction, the “blob catalogs” represent the underlying astrophysical signal and thus allowed us to assess the effects of noise, transfer function, and data-reduction–induced biases.

\begin{figure}[!t]
\centering
\includegraphics[width=1\columnwidth]{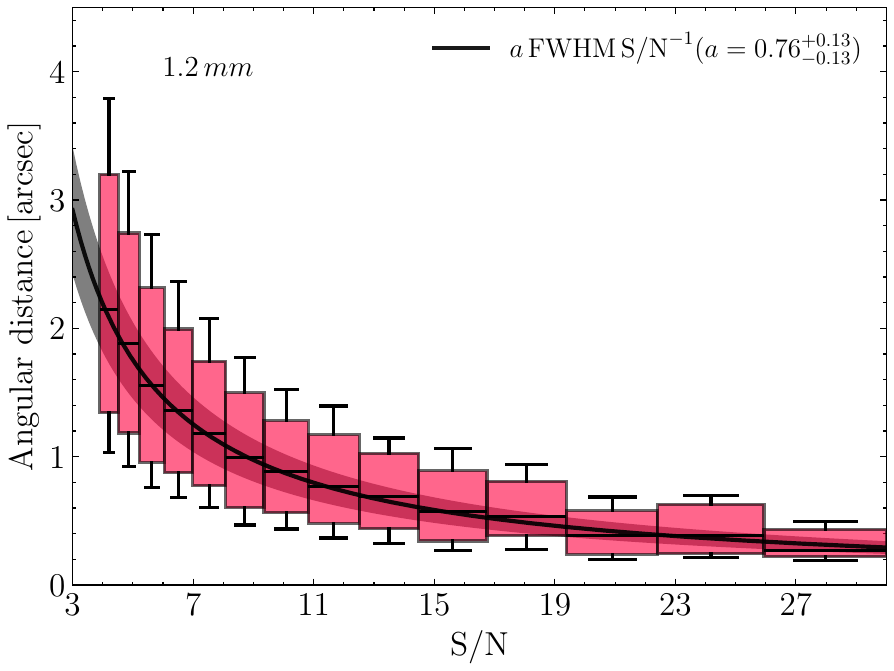}
\includegraphics[width=1\columnwidth]{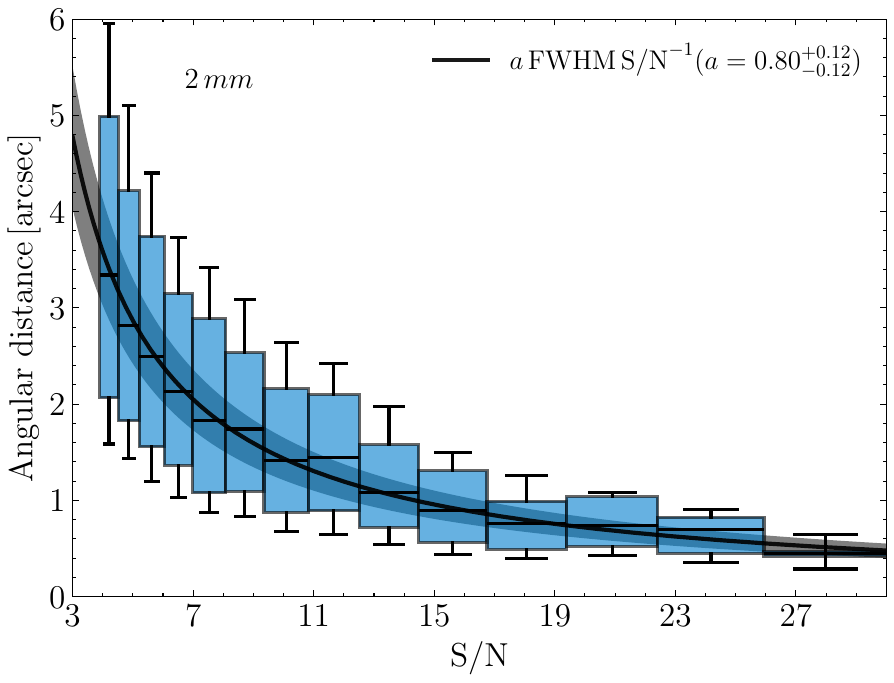}
\caption{Angular separation ($\Delta r$) between the true source positions in the blob catalog (noise-free map) and the recovered positions in the realistic simulated catalog (simulated map) from SIDES as a function of S/N at 1.2\,mm (\textit{top}) and 2\,mm (\textit{bottom}). For each S/N bin, the box shows the interquartile range (25th--75th percentiles), while the whiskers indicate the 16th--84th percentiles of the distribution. The width of the box corresponds to the S/N bin size, and the horizontal black line marks the median $\Delta r$. The solid black line shows the best-fit relation $\Delta r = a\,{\rm FWHM}/{\rm (S/N)}$, and the shaded region represents the $1\sigma$ confidence interval obtained using Markov chain Monte Carlo (MCMC).}
\label{fig: position_blob_simulated}
\end{figure}

We compared the positions of the sources detected in the blob catalog and in the simulated catalog (i.e., source position in the noise-free map versus source position in the simulated map). For that purpose, we cross-matched the source positions using a matching radius of $\rm 0.5\times FWHM$, and we selected the closest source in case of multiple matches. For each matched pair, we computed the angular separation $r$ between the true positions from the blob catalog and those recovered in the simulated catalog. We show on Fig.\,\ref{fig: position_blob_simulated} the resulting angular separation as a function of S/N. As expected \citep{Condon+97, Ivison+07}, $\Delta r$ follows $\Delta r = a\times{\rm FWHM}/{\rm (S/N)}$ at both 1.2 and 2\,mm, with $a=0.76$ and $a=0.80$, respectively, in good agreement with the theoretical expectation of $1/\sqrt{2\ln 2}$.

\subsubsection{Purity}\label{sec: purity}
\begin{figure*}[!t]
\centering
\includegraphics[width=0.48\linewidth]{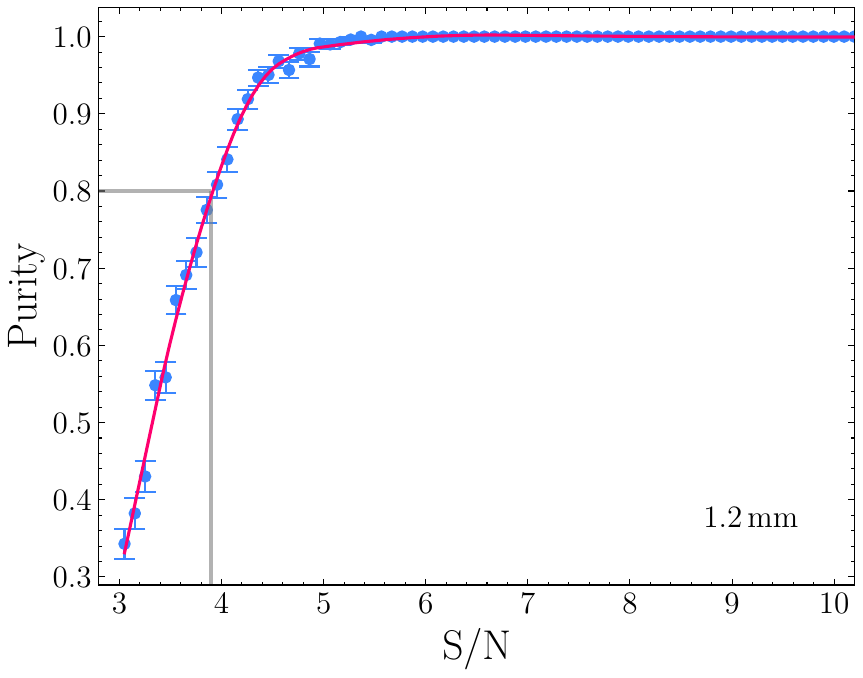}
\includegraphics[width=0.48\linewidth]{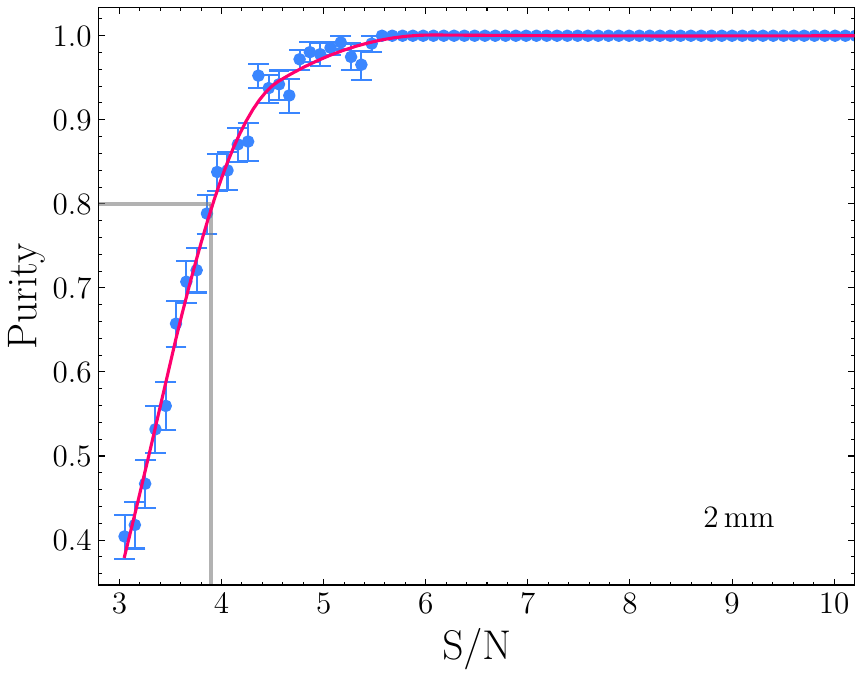}
\caption{Purity of detected sources at different S/N at 1.2\,mm (\textit{left}) and 2\,mm (\textit{right}), estimated from the simulations (see Sect.~\ref{sec: purity}). The red line shows the fit by a spline function to the results. The solid gray line indicates the S/N cut of 3.9, corresponding to 80\% purity.}
\label{fig: purity}
\end{figure*}

A key quantity for characterizing our catalog is its purity, defined as the probability that an extracted source is real, or equivalently, one minus the fraction of spurious sources. To estimate the purity, we used the ``simulated catalogs'' derived from 117 simulated observations. These ``simulated catalogs'' contains both true and spurious sources, or true and false positive (TP+FP). We also built ``null catalogs'' from the half-difference maps, following the process described in Sect.~\ref{subsec: data_redu}. 
Since the astrophysical signal is removed by construction, these catalogs contain only spurious sources and therefore yield only the false-positive counts (FP).
We computed the true positive detection (TP) as the count difference between the ``simulated catalogs'' and ``null catalogs,'' and we defined the purity as
\begin{equation}
    P = \frac{TP}{TP + FP}
    \label{eq: purity}
.\end{equation}
This methodology enabled us to estimate the purity as a function of S/N, as shown in Fig.~\ref{fig: purity}, where the results are fitted with a spline function. The purity reaches $80\%$ at 3.9 S/N for 1.2\,mm and 2\,mm, and $>95\%$ at S/N $>4.5$ for 1.2\,mm and S/N $>4.6$ for 2\,mm. This procedure allows us to establish a detection threshold at S/N $= 3.9$ ($80\%$ purity) that we use in subsequent analyses.
This procedure differs from that presented in \citetalias{Bing+23} to estimate the purity (their Sect.~3.1), where a match was made to the input blob catalog.

\subsubsection{Flux correction}\label{sec: flux_correction}

The raw flux densities measured in our catalog are affected by the data reduction process (pipeline transfer function) and noise. During data reduction, a fraction of the source flux can be filtered out when sky noise and polynomial baselines are removed. Furthermore, because these catalogs are constructed using S/N as the detection criterion, noise can bias the detection toward sources that coincide with positive noise fluctuations. This bias leads to an overestimate of the flux for faint sources just above the threshold and produces the well-known ``flux boosting.''

In this work, we do not separately measure the impact of noise and transfer function. Instead, we estimated the effective ratio between the input flux, measured in the blob maps, and the flux extracted from the simulated observations as a function of S/N, following \citetalias{Bing+23}. From now on, we refer to the deviation between the input and output flux (i.e., the output-to-input flux ratio), which is influenced by both effects, as the effective flux boosting.

To estimate the effective flux boosting, we performed cross-matching between the input blob catalog and the simulated catalog extracted from the same simulated observation. We consider an input blob source to be recovered if a source in the simulated catalog is detected within $0.5\times$FWHM of the input position, above the S/N threshold set in Sect.~\ref{sec: purity}. In cases with multiple matches within the matching radius, the closest source was selected.  We reduced the matching radius compared to \citetalias{Bing+23} (from $0.75\times$FWHM to $0.5\times$FWHM) in order to minimize potential false detections, taking into account that our data are deeper.

\begin{figure*}[!t]
\centering
\includegraphics[width=0.48\linewidth]{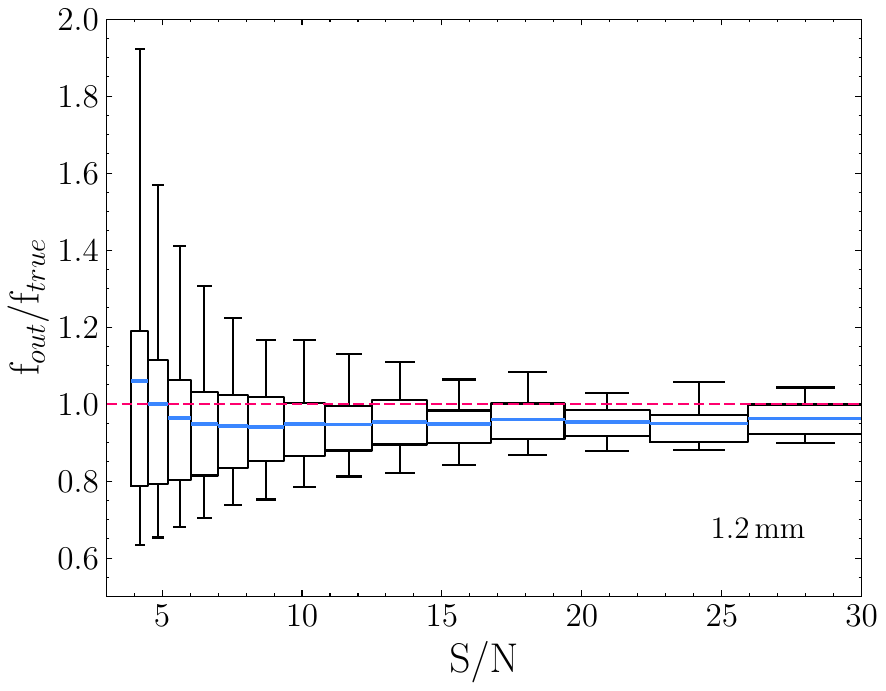}
\includegraphics[width=0.48\linewidth]{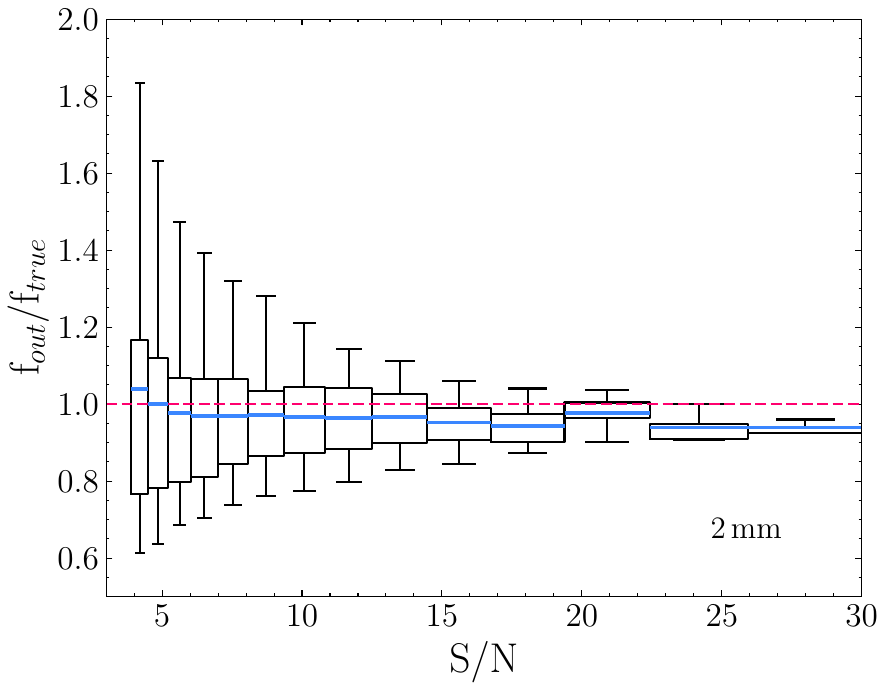}
\caption{Effective flux boosting ratio as a function of S/N at 1.2\,mm (\textit{left}) and 2\,mm (\textit{right}). The flux correction is estimated as the ratio between the measured flux from the simulated observations ($f_{out}$) and the input flux from the blob catalog ($f_{true}$). The box at each S/N represents the effective flux boosting distribution enclosed between the 25th and 75th percentiles, while the upper and lower bounds show the 5th and 95th percentiles of the distribution. The width of each box shows the width of the S/N bin. The blue lines show the mean flux boosting, and the dotted red line shows the unity effective flux boosting. }
\label{fig: flux_correction_mean}
\end{figure*}

In Fig.~\ref{fig: flux_correction_mean}, we present the distributions of the effective flux boosting for the two bands as a function of S/N. The mean effective flux boosting drops below unity for both bands at high S/N ($\lesssim 0.94$), indicating that the flux filtering is stronger than the flux boosting. However, for lower S/N ($<5.2$), the flux boosting compensates the effect of the transfer function, slightly exceeding it at S/N $<4.5$. This behavior diverges from that reported in \citetalias{Bing+23}, where the mean effective flux boosting is above unity for the COSMOS field.  However, this difference is reasonable given that our observations are deeper; consequently, the flux boosting is less significant than the filtering effect.  
This behavior was also observed in their results for the GOODS-N field, where the observations are approximately three times deeper than ours.
The interquartile range of the distributions of the effective flux boosting is as high as $\sim38\%$ at S/N $\sim 4$, and drops to less than $10\%$ at S/N $>14$. Consequently, the scatter in $f_{\rm out}/f_{\rm true}$ is reduced compared to \citetalias{Bing+23}. However, the uncertainties at low S/N remain large, as indicated by the error bars, which encompass the 5-95\% range of the data. We account for this effect in our flux correction.

\section{N2CLS source catalogs}\label{sec: N2CLS_catalog}

We constructed blind source catalogs at 1.2\,mm and 2\,mm following the method described in Sect.~\ref{subsec: source_extraction}, to detect the sources and measure their fluxes independently in each band. We used an S/N threshold as the only detection parameter, set to achieve 80\% purity (S/N>3.9 for both bands). The raw fluxes obtained from the PSF fitting (S$_\mathrm{raw}$) were corrected using the effective flux boosting (Sect.~\ref{sec: flux_correction}) to derive the corrected flux density for each source (S$_\mathrm{corr}$).

To apply the effective flux boosting and correct the flux density of each source, we implemented a Monte Carlo scheme that incorporates the boosting distributions (see Fig.~\ref{fig: flux_correction_mean}) as well as the measured flux density errors. For each S/N bin, the effective flux boosting was modeled with a log-normal distribution. We then generated 10,000 random realizations of the effective flux boosting drawn from the distribution at each bin. These realizations were applied to each source according to its S/N detection. 
Additionally, we incorporated the uncertainty in the sources’ raw flux by modeling each source’s observed flux density as a Gaussian distribution centered on  S$_\mathrm{raw}$, with $\sigma$ corresponding to the measured flux uncertainty.
For each Monte Carlo realization, a flux value was randomly drawn from the Gaussian distribution and then corrected using one realization from the sampled flux-boosting distribution. In this way, the Monte Carlo simulations produced a new corrected flux density distribution for each source, propagating both the uncertainty in the effective flux boosting and the measured uncertainty in the raw flux.
For sources detected at S/N$\sim 5$, the flux amplitude correction is typically $\sim$5\% for the median flux, while the associated uncertainty can increase by 
$\sim$60-90\%, mainly due to the scatter in the effective flux boosting (Fig.~\ref{fig: flux_correction_mean}).
Consequently, the corrected flux uncertainties are larger than those implied by the detection S/N, reflecting the combined effects of noise and the scatter in the effective flux-boosting distributions.
Note that the absolute calibration uncertainties are not included in the flux error bars. These uncertainties have two components: a systematic calibration uncertainty of 5\%, arising from the planet models used to calibrate the NIKA2 data, and the point-source rms calibration uncertainties, which are about 6\% at 1.2\,mm and about 3\% at 2\,mm \citep{Perotto2020}.\\

The total number of sources blindly detected with $\geq$80\% purity are 301 at 1.2\,mm and 124 at 2\,mm. Compared to \citetalias{Bing+23}, we report an increase of approximately 54\% in the number of sources detected at 1.2\,mm, and 63\% at 2\,mm. We detect sources as faint as $1.14_{-0.41}^{+0.55}$\,mJy and $0.34_{-0.11}^{+0.14}$\,mJy, and as bright as $9.20_{-0.64}^{+0.52}$\,mJy and $2.07_{-0.12}^{+0.13}$\,mJy, at 1.2\,mm and 2\,mm respectively. The median corrected flux density is $\sim2.28$\,mJy at 1.2\,mm and $\sim0.62$\,mJy at 2\,mm.\\

We built an N2CLS master catalog by cross-matching the blind catalogs extracted at 1.2\,mm and 2\,mm using the matching radius: 
\begin{equation}
    r_{\rm matching} = \sqrt{\rm (FWHM_1/2)^2 + (FWHM_2/2)^2},
    \label{eq: matching_radius}
\end{equation}
where FWHM$_1$=11.6$\arcsec$ and FWHM$_2$=18$\arcsec$ are the beam sizes of the 1.2\,mm and 2\,mm maps, respectively. 
The master catalog comprises 323 sources with S/N $\geq 3.9$ ($\geq 80\%$ purity), of which 104 appear in both bands, 197 are detected only at 1.2\,mm, and 22 only at 2\,mm.
We identified nine N2CLS sources with separations $r > 6.5\arcsec$ between the 1.2 and 2\,mm positions; these are discussed in Appendix\,\ref{app:master_sep}. 
The 22 sources detected only at 2\,mm are unlikely to be DSFGs. Their non-detection at 1.2\,mm implies $S_{\rm 2\,mm}/S_{\rm 1.2\,mm}$ colors substantially redder than those of the dust-dominated population, which is most likely explained by a contribution from synchrotron emission, rising toward longer wavelengths \citep{Bethermin+26}.
Indeed, 10 of the 16 identified sources have a radio counterpart (see Sect.~\ref{sec: proxy_identification}), and their median redshift ($z\sim1.6$) is well below that of the 1.2\,mm or 2\,mm high-purity sample ($z=2.8$ and $z=3$, respectively; see Sect.~\ref{sec:n2cls_galaxies}), indicating that they trace a different, lower-redshift radio-galaxy population.
The 1.2+2\,mm cross-matched catalog constitutes our master N2CO catalog (Table~\ref{tab: N2CLS_master}), serving as the starting point for subsequent analysis. For brevity, only the first few entries are shown in Table~\ref{tab: N2CLS_master}, and the full catalog is publicly available at the CDS and on our data release page (\url{https://data.lam.fr/n2cls}). 

\begin{table*}
\centering
\caption{N2CLS source master catalog (1.2+2\,mm) in the COSMOS field.}
\label{tab: N2CLS_master}
\begin{tabular}{l l c c c c c c c}
\hline
\hline
Name & N2CLS name & R.A.  & Dec & $\mathrm{S/N}_{1.2\,\mathrm{mm}}$ & $\mathrm{S/N}_{2\,\mathrm{mm}}$ & $S_{1.2\,\mathrm{mm}}^{\mathrm{corr}}$ & $S_{2\,\mathrm{mm}}^{\mathrm{corr}}$ \\
 &  & deg  & deg &  &  & mJy & mJy \\
\hline
N2CO J100008+022613 & N2CO\_1\_1 & 150.033523 & 2.436845 & 28.12 & 22.04 & $9.20_{-0.64}^{+0.52}$ & $2.07_{-0.12}^{+0.13}$ \\
N2CO J100016+021549 & N2CO\_1\_2 & 150.065030 & 2.263624 & 22.61 & 17.07 & $7.18_{-0.58}^{+0.48}$ & $1.62_{-0.13}^{+0.13}$ \\
N2CO J095943+022938 & N2CO\_1\_3 & 149.928411 & 2.493874 & 22.42 & 14.35 & $7.71_{-0.62}^{+0.50}$ & $1.38_{-0.16}^{+0.16}$ \\
N2CO J100057+022014 & N2CO\_1\_4 & 150.237987 & 2.337248 & 21.82 & 15.97 & $6.84_{-0.45}^{+0.46}$ & $1.46_{-0.13}^{+0.15}$ \\
N2CO J095957+022731 & N2CO\_1\_5 & 149.988669 & 2.458708 & 20.44 & 16.06 & $6.87_{-0.44}^{+0.48}$ & $1.55_{-0.14}^{+0.15}$ \\
N2CO J100123+022006 & N2CO\_1\_6 & 150.345515 & 2.334991 & 17.67 & 11.89 & $7.18_{-0.65}^{+0.63}$ & $1.49_{-0.20}^{+0.22}$ \\
N2CO J100020+023205 & N2CO\_1\_7 & 150.082192 & 2.534661 & 17.00 & 9.44 & $5.49_{-0.49}^{+0.49}$ & $0.87_{-0.14}^{+0.17}$ \\
N2CO J100000+023442 & N2CO\_1\_8 & 149.997104 & 2.578407 & 15.16 & 12.04 & $5.71_{-0.61}^{+0.61}$ & $1.42_{-0.20}^{+0.21}$ \\
N2CO J100029+023204 & N2CO\_1\_9 & 150.119494 & 2.534457 & 14.23 & 9.17 & $4.66_{-0.52}^{+0.54}$ & $0.88_{-0.16}^{+0.17}$ \\
N2CO J100032+021242 & N2CO\_1\_10 & 150.132477 & 2.211578 & 13.98 & 8.68 & $4.36_{-0.50}^{+0.52}$ & $0.81_{-0.15}^{+0.16}$ \\
... & ... & ... & ... & ... & ... & ... & ... \\
\hline
\hline
\end{tabular}
\tablefoot{The R.A. and Dec columns provide the coordinates of the N2CLS sources, measured at 1.2\,mm or at 2\,mm when no 1.2\,mm detection is available. The columns $\mathrm{S/N}_{1.2\,\mathrm{mm}}$, $\mathrm{S/N}_{2\,\mathrm{mm}}$, $S_{1.2\,\mathrm{mm}}^{\mathrm{corr}}$, and $S_{2\,\mathrm{mm}}^{\mathrm{corr}}$ list the S/N and the corrected flux densities at 1.2\,mm and 2\,mm, respectively. The catalog is sorted by $\mathrm{S/N}_{1.2\,\mathrm{mm}}$. When the source was not detected at 1.2\,mm, sorting was done by $\mathrm{S/N}_{2\,\mathrm{mm}}$. 
The full catalog is publicly available at the CDS and on our data release page (\url{https://data.lam.fr/n2cls}).
}
\end{table*}

\section{Multiwavelength source identification}\label{sec: multiwave_ident}
N2CLS represents the largest-area blind survey at this depth at millimeter wavelengths with a single-dish telescope, designed to uncover high-redshift dusty galaxies \citep{Bing+23,Berta+25,Bethermin+26, lagache2026, Ponthieu2026}.
However, the relatively low angular resolution of NIKA2 at 1.2\,mm and 2\,mm makes it challenging to unambiguously associate our N2CLS galaxies with counterparts in the extensive ancillary dataset available for COSMOS. This issue is well known in (sub)millimeter astronomy, where single-dish surveys often suffer from source confusion and positional uncertainty. In this section we outline the identification strategy adopted to locate reliable counterparts and to then construct the UV-to-radio spectral energy distributions (SEDs) of the N2CO sources. This strategy follows the approach previously developed for the N2CLS galaxies in the GOODS-N field, as described in \cite{Berta+25}.

\subsection{Source identification}\label{sec: proxy_identification}
A robust method to unambiguously pinpoint the correct counterparts of N2CLS galaxies is to rely on interferometric follow-up observations. Such approaches have been successfully implemented using the submillimeter array \citep[e.g.,][]{Cowie+17} or ALMA \citep[e.g.,][]{Stach+19} for SCUBA2 sources, and NOEMA \citep[e.g.,][]{Berta+25} for N2CLS sources in the GOODS-N field.\\
As detailed in \cite{Berta+25}, we adopt this approach to determine the precise positions of our N2CLS sources and to identify their multiwavelength counterparts. Our primary positional reference is provided by the high-angular resolution (sub)millimeter observations from ALMA and NOEMA.
For the remaining sources, we utilized radio observations, leveraging the well-established Far IR-radio correlation in star-forming galaxies \citep[e.g.,][]{Smail+00, Ivison+02, Borys+04}. Radio counterparts, however, should be treated with caution, as the positive $k$-correction at these wavelengths tends to favor lower-redshift systems, which can result in incorrect identifications \citep[e.g.,][]{Chapman+05}.
To obtain the most accurate positions for our galaxies, we therefore searched for counterparts in the available catalogs following the order of priority below:

\paragraph{AS2COSMOS} \citep{AS2COSMOS}:  This catalog is based on ALMA Band-7 follow-up observations of the 160 brightest 850\,$\mu$m sources ($\rm S_{850\mu m}\gtrsim8$\,mJy) in the \textsc{main} region of the SCUBA2-COSMOS survey \citep[S2COSMOS;][]{Simpson+19}. This region reaches a median sensitivity of 1.2\,mJy/beam over the 1.6 square degrees. The observations were carried out using a standard continuum correlator setup, with four basebands providing a total bandwidth of 7.5\,GHz centered at 343\,GHz (870\,$\mu$m). The final sample includes 24 more sources that were identified in the archival ALMA Band-7 imaging data. The spatial resolution of the ALMA observations present small variations (FWHM$\sim$0.8\arcsec), and we used a conservative beam size of 1\arcsec\ for the cross-matching between our N2CO master catalog and the AS2COSMOS catalog (following Eq.\,\ref{eq: matching_radius}). 
Including some ALMA archival data, the full observed sample comprises 182 SCUBA2 sources with deboosted/deblended flux densities $\rm S_{850\mu m}>6.2$\,mJy. They are resolved into 260 individual galaxies detected in the ALMA maps, highlighting multiple detections within single SCUBA2 sources. In our analysis, we identify 73 counterparts in the AS2COSMOS catalog, for 53 N2CLS sources.

\paragraph{N2CLS NOEMA follow-up} (W23CJ; PI: A. Beelen):
An ongoing NOEMA program targets 22 N2CLS sources, building on the successful identification strategy previously applied to the GOODS-N field. The observational design utilizes track-sharing mode in the C configuration across two bands to match the NIKA2 frequencies. By achieving an angular resolution of approximately 1.1-1.4\arcsec, these observations facilitate the accurate cross-matching of NIKA2 detections with deep ancillary datasets.
At the time of writing, 16 sources have been observed, including the source N2CO\_1\_173 (S/N$\sim$5.26) which remains unidentified. 
The calibrated visibility tables from the lower (LSB) and upper (USB) sidebands were collapsed into a single double-sideband (DSB) continuum dataset. We performed a robust multi-component uv fit on this combined continuum data to accurately localize the sources within the observed field of view. High-resolution continuum images were also produced through standard deconvolution and cleaning procedures.
These observations allowed us to identify 18 new counterparts for 16 N2CLS sources, including three pairs of multiple counterparts, for N2CO\_1\_74, N2CO\_1\_86, and N2CO\_1\_241-N2CO\_2\_76. The latter represents a special case, where the follow-up of N2CO\_1\_241 revealed the presence of a second counterpart. This additional proxy is associated with N2CO\_2\_76, a 2\,mm-only source located 16\arcsec\ from N2CO\_1\_241 and therefore blended with it.

\paragraph{A3COSMOS} \citep{A3COSMOS}:  This survey is built from all available public (sub)millimeter ALMA observations, retrieved from the archive using automated mining pipelines. The procedure involves data reduction and continuum imaging, followed by two photometric extraction approaches: a blind source extraction and a prior-position flux measurement based on a multiwavelength catalog compilation (optical, near-infrared, mid-infrared, and radio).
We are using the latest available version of the catalogs,\footnote{Data release version 20250312: \url{https://sites.google.com/view/a3cosmos}.} which contain $4287$ pointings from 244 projects. The prior catalog may include sources with multiple observations (S/N $>$ 4.35), which can result in positional offsets of up to 0.7\arcsec. To address this, we built a master prior catalog by computing an S/N-weighted average of the flux-extracted positions. The blindly extracted catalog contains sources with S/N$>$5.4. Since some sources may be detected multiple times (e.g., in  different programs and/or at different wavelengths), we first grouped the detections within 1\arcsec, then built a master blind catalog in the same way as for the prior catalog. For our analysis, and to complement the master prior catalog, we used the unique sources extracted from the master blind catalog and not present in the prior catalog. As for AS2COSMOS, we used 1\arcsec\ as the beam size for the cross-matching with our master N2CO catalog. We identify 133 counterparts in the master prior catalog and four counterparts in the master blind catalog for 118 N2CLS sources, plus one N2CLS source (N2CO\_1\_4) that shares other proxies from AS2COSMOS \citep[the proxy from A3COSMOS is at a different position; this N2CLS source corresponds to the proto-cluster at z$\sim$2.5, ][]{Stott+16}.

\paragraph{CHAMPS} (ALMA 2023.1.00180.L; PI: A. Faisst): This is an ALMA 0.18\,deg$^2$ blank-field survey at 1.2\,mm covering the JWST MIRI footprint \citep{Faisst2026,Zavala2026}. As of November 2025, the CHAMPS team provided a preliminary catalog defined using a 5$\sigma$ detection threshold, corresponding to an estimated $\sim$95\% purity \citep{Martinez2026}. This catalog is used solely as a positional reference; its flux measurements are not employed in our analysis. Following the approach adopted for A3COSMOS, where many ALMA pointings are combined, we grouped sources separated by less than 1\arcsec, resulting in a catalog of 839 sources with a typical synthesized beam size of $\sim$2\arcsec\ (major axis). 
Cross-matching with N2CLS allows this catalog to serve as a positional reference for 51 N2CLS sources, yielding 54 counterparts.

\paragraph{Radio}: The radio proxies are drawn from the Very Large Array (VLA)-COSMOS 3\,GHz Large Project \citep{3GHZVLA}, which covers 2\,deg$^{2}$. The final mosaic reaches a median rms of 2.3\,$\mu$Jy\,beam$^{-1}$ at an angular resolution of 0.75\arcsec. We identified 50 radio counterparts associated with 46 N2CLS sources. In addition, one N2CLS source (N2CO\_1\_67) has an additional proxy in A3COSMOS at a different position.\\

We visually inspected all our sources to verify the accuracy of the proxy identification and to check for any missing proxies. During this process, we found four MeerKAT sources \citep[from the MIGHTEE - MeerKAT International Gigahertz Tiered Extragalactic Exploration - survey;][]{MeerKat}, with JWST counterpart,  for four N2CLS sources, plus one VLA-COSMOS 1.4\,GHz source \citep{1.4GHzVLA} for one N2CLS galaxy. In addition, we discarded 11 identifications for the following reasons: multiple proxies correspond to the same source (seven from the A3COSMOS prior catalog), the proxy lies at the very edge of the NIKA2 beam (one VLA counterpart), or the proxy likely represents a spurious low-S/N detection (three from the A3COSMOS prior catalog). Those 11 proxies have already been removed from the numbers above.
This leaves 34 N2CLS sources out of 323 without any proxy: three detected at both 1.2\,mm and 2\,mm, 25 detected only at 1.2\,mm, and six detected only at 2\,mm.\\

\begin{figure}
\centering
\includegraphics[width=1\columnwidth]{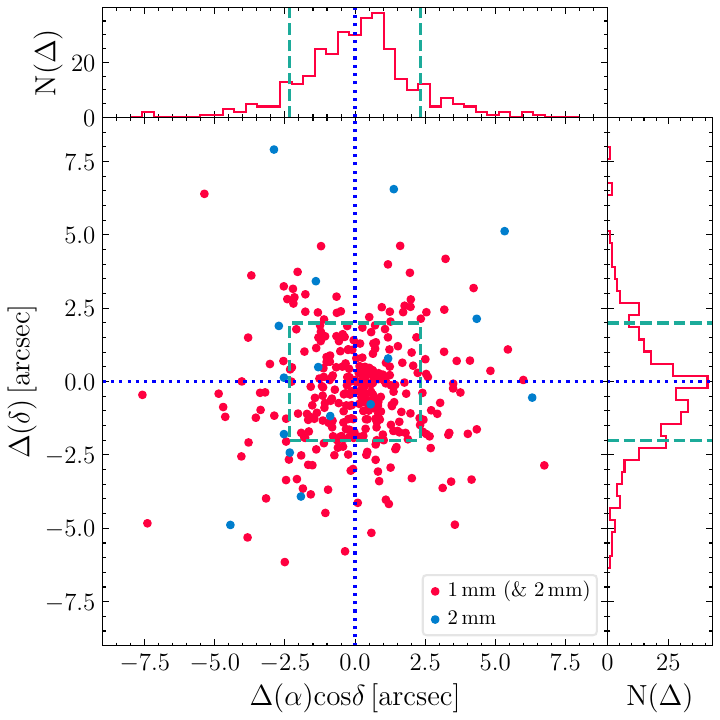}
\caption{Angular distance between the coordinates of the N2CO sources and their identified proxy (Sect.~\ref{sec: proxy_identification}). The red points show the 1.2\, and 1.2+2\,mm N2CLS sources, while the blue points show the 2\,mm-only sources in the N2CO master catalog. The red histograms show the total offset distribution. The dotted blue line marks zero offset, while the dashed green box and lines indicate the rms dispersion of the total offset distribution.}
\label{fig: proxy_offset}
\end{figure}

\begin{table*}[!t]
\centering
\caption{Proxy identification of the N2CO master catalog for $\geq$95\% purity (S/N$\geq$4.5 at 1.2\,mm, S/N$\geq$4.6 at 2\,mm).}
\resizebox{\linewidth}{!}{
\begin{tabular}{cccccccccccccc}
                \cline{3-12}
            & &  \multicolumn{8}{c}{{Proxy Catalog}} & &\\
                \cline{3-12}
               & N2CLS Sources & 
               \multicolumn{2}{c}{AS2COSMOS}
               & \multicolumn{2}{c}{{A3COSMOS}} 
               & \multicolumn{2}{c}{{NOEMA}} 
               & \multicolumn{2}{c}{{CHAMPS}} 
               & \multicolumn{2}{c}{{Radio}} 
               & No proxy
               & \multirow{2}{*}{\begin{tabular}[c]{@{}c@{}} Total N. \\ of proxies \end{tabular}}  \\ 
            &  & Single & Multi & Single & Multi & Single & Multi & Single & Multi & Single & Multi & & \\ 
                \cline{2-14}    
                
1.2\,mm~\&~2\,mm$\rm^a$  & 101   &  30   & 15    & 29    & 4(+1)$\rm^b$  & 4     & 2     & 8  & -- & 6  & (+1)$\rm^c$               & 3     & 124  \\
1.2\,mm           & 132   &  6    & 1     & 48    & 8             & 8     & --    & 24 & 3  & 21  & 2         & 11    & 137 \\
2\,mm             & 10    &  --   & --    & --    & --            & 1$\rm^d$     & --    & 2  & -- & 5   & --            & 2     & 8  \\ \hline
TOTAL & 243      &  36   & 16    & 77    & 12(+1)        & 13    & 2     & 34 & 3  & 32  & 2(+1)         & 16    & 269 \\ \hline
\end{tabular}
}
\tablefoot{Proxy identifications are based on (sub)millimeter and radio catalogs from AS2COSMOS, A3COSMOS, NOEMA, CHAMPS, VLA 3\,GHz, VLA 1.4\,GHz, and MeerKAT. The table reports the number of N2CLS sources with single or multiple proxy associations in each dataset, the sources without an identified proxy, and the total number of proxies.
$\rm^{a}$ Sources detected at 1.2\,mm with $\geq$95\% purity and with a counterpart at 2\,mm (S/N$>3.9$), not necessarily at high purity. Among these 101 sources, 80 are detected at 2\,mm with $\geq$95\% purity. The total number of 2\,mm sources detected at $\geq$95\% purity is 90 (80 with a 1.2\,mm counterpart and ten detected only at 2\,mm). 
$\rm^{b,c}$ Proxies from different catalogs added during the manual identification process (see Sect.~\ref{sec: proxy_identification}) that are associated with the same N2CO source. These entries are listed in the table but are not included in the total source count.
$\rm^b$ Corresponds to a counterpart in A$3$COSMOS (N2CO\_1\_4\_e) sharing the same N2CLS source with four AS2COSMOS counterparts. $\rm^c$ A proxy from the VLA-3\,GHz catalog (N2CO\_1\_67\_c), which shares the same N2CLS source as two A3COSMOS counterparts (N2CO\_1\_67\_a and N2CO\_1\_67\_b).
$\rm^{d}$ It corresponds to the counterpart of N2CO\_2\_76, identified in the NOEMA follow-up of N2CO\_1\_241 as a secondary proxy (see Sect.~\ref{sec: proxy_identification}). It is not classified as a multiple association, as the resulting proxies are linked to different N2CLS sources.}
\label{tab: identification_95_purity}
\end{table*}

\begin{table*}[!t]
\centering
\caption{Same as Table~\ref{tab: identification_95_purity} but for N2CO sources with purity between 80\% and 95\%, i.e., 3.9$<$S/N$<$4.5(4.6) at 1.2(2)\,mm.}
\resizebox{\linewidth}{!}{
\begin{tabular}{cccccccccccccc}
                \cline{3-12}
            & &  \multicolumn{8}{c}{{Proxy Catalog}} & &\\
                \cline{3-12}
               & N2CLS Sources & 
               \multicolumn{2}{c}{AS2COSMOS}
               & \multicolumn{2}{c}{{A3COSMOS}} 
               & \multicolumn{2}{c}{{NOEMA}} 
               & \multicolumn{2}{c}{{CHAMPS}} 
               & \multicolumn{2}{c}{{Radio}} 
               & No proxy
               & \multirow{2}{*}{\begin{tabular}[c]{@{}c@{}} Total N. \\ of proxies \end{tabular}}  \\ 
            &  & Single & Multi & Single & Multi & Single & Multi & Single & Multi & Single & Multi & & \\ 
                \cline{2-14}    
                
1.2\,mm~\&~2\,mm$\rm^{a}$  & 3   &  --   & --    & 2     & 1 & --    & --    & -- & -- & --  & -- & --    & 4 \\
1.2\,mm         & 65    &  1    & --    & 19    & 4 & 1    & --    & 14  & -- & 12  & -- & 14    & 55 \\
2\,mm           & 12    &  --   & --    & 3     & -- & --    & --    & -- & -- & 4   & 1  & 4    & 9 \\ \hline
TOTAL & 80     &  1    & --   & 24    & 5 & 1     & --     & 14  & --  & 16  & 1  & 18   & 68 \\ \hline
\end{tabular}
}
\tablefoot{$\rm^{a}$ Sources detected at 1.2\,mm with purity between 80\% and 95\%, and with a counterpart at 2\,mm. The total number of 2\,mm sources detected at purity between 80\% and 95\% is 34 (22 with a 1.2\,mm counterpart, plus 12 detected only at 2\,mm).}
\label{tab: identification_80_purity}
\end{table*}

Table~\ref{tab: identification_95_purity} presents a summary of the proxies used for source identification and their corresponding results. For the 95\% purity sample (243 sources), 193 are identified through ALMA+NOEMA and 34 from the radio data. 16 remain not identified, which represents $\sim$7\% of the sample (see Sect.\,\ref{sect:unidentified}). At this purity level, we identified a single proxy for 192 N2CLS sources and multiple proxies for 35 sources. For the 80 N2CLS sources with purity between 80 and 95\% (Table~\ref{tab: identification_80_purity}), 45 are identified through ALMA+NOEMA, 17 from the radio, and 18 remain unidentified ($\sim$22\%). At this purity level, 56 N2CLS sources exhibit a single proxy, while 6 sources show multiple components. In Fig.~\ref{fig: proxy_offset}, we show the angular distance distribution between the N2CLS coordinates and the proxy position. The median separation for the 337 identified proxies is 2\arcsec. Among them, 24 have a separation distance higher than 5\arcsec: six correspond to a single counterpart (four identified with the radio, one with ALMA and one with NOEMA), while 18 are multiple counterparts (more than one proxy per N2CLS source), including 14 identified with ALMA (with three belonging to the proto-cluster at z$\sim$2.5 that contains five proxies in total) and four identified with the radio.

\subsection{Multiwavelength counterparts}\label{sec: multi_counterpart}

Once a proxy had been identified for each N2CLS source, we searched for its multiwavelength counterparts to construct its SED. We cross-matched the proxy positions with the ancillary catalogs using Eq.~\ref{eq: matching_radius}. In this equation, $\mathrm{FWHM}_1$ corresponds to the beam size of the observations used to determine the proxy position (e.g., NOEMA), while $\mathrm{FWHM}_2$ corresponds to the beam size of the observations at the other wavelengths. When multiple counterparts were found, we selected the one closest to the proxy position.

COSMOS is one of the most extensively studied extragalactic fields, with broad multiwavelength coverage. The field was first observed in 2003 with the Advanced Camera for Surveys (ACS) on board the \textit{Hubble} Space Telescope (HST), covering an area of $\sim2$\,deg$^2$. Since then, COSMOS has been observed by nearly all major space- and ground-based observatories, including \textit{Chandra}, \textit{XMM-Newton}, GALEX, \textit{Spitzer}, \textit{Herschel}, the Very Large Array (VLA), the Subaru Telescope, and the JWST, among others. These observations have produced extensive photometric and spectroscopic datasets. To construct the multiwavelength SEDs of the N2CLS sources, we compiled publicly available data covering the optical to radio wavelengths. The catalogs used are briefly described below.

\subsubsection{Optical and near-infrared ancillary data}

\paragraph{COSMOSWeb:} We use the latest COSMOS catalog\footnote{\url{https://cosmos2025.iap.fr/}.} compilation \citep{shuntov+25}, which include deep JWST/NIRCam imaging in the F115W, F150W, F277W, and F444W bands, as well as MIRI F770W observations, covering $0.54 \text{ deg}^2$ and $0.2 \text{ deg}^2$, respectively. 
This catalog provides photometry, morphological measurements, and photometric redshifts for more than 700,000 galaxies. The JWST imaging is combined with 32-band photometry spanning 0.3–8.0\,$\mu$m, incorporating extensive ancillary data (from e.g., HST, VISTA). Source detection was performed using a multi-band $\chi^{2}$ map constructed from the NIRCam bands after PSF homogenization to the F444W PSF.
The NIRCam bands reach a 5$\sigma$ depth of 27.2\,mag for F115W, 27.4\,mag for F150W, 28.1\,mag for F277W and 28.0\,mag for F444W. F770W (MIRI) band reaches 25.2\,mag. 
Photometric redshifts were estimated using the \texttt{LePHARE} code \citep{Ilbert2006}, achieving an accuracy of $\sigma_{\mathrm{MAD}} = 0.012$ for sources with $m_{\mathrm{F444W}} < 28$ and $\sigma_{\mathrm{MAD}} \lesssim 0.03$ for the full sample, as assessed through comparison with spectroscopic redshifts up to $z \sim 9$. 

\paragraph{COSMOS2020:} We also used the COSMOS2020 catalog \citep{COSMOS2020} for sources that are not covered by COSMOSWeb and for the photometric redshifts (see Sect.~\ref{sec:z_distrib} and Appendix\,\ref{app: photo_redshift}). This release covers $2~\text{deg}^2$ of the COSMOS field and represents the most comprehensive multiwavelength compilation  prior to the JWST. The catalog provides photometric measurements for approximately 1.7 million sources. The photometry spans roughly 40 bands covering the near-ultraviolet to the mid-infrared (0.3–4.5\,$\mu$m).
Source detection was performed on a combined $\chi^{2}$ detection map constructed from the $i z Y J H K_s$ bands.
The catalog reaches $3\sigma$ depths of approximately $27.0$\,mag in the HSC $i$-band and $25.2$\,mag in the VISTA $K_s$-band. The photometry was extracted using two independent approaches: the ``Classic'' catalog, which employs \texttt{SExtractor} aperture photometry on PSF-homogenized images, and the ``Farmer'' catalog, which utilizes \texttt{The Tractor} for parametric profile-fitting to account for source blending.
The photometric redshifts ($z_{\text{phot}}$) were estimated using the \texttt{LePHARE} \citep{Ilbert2006} and \texttt{EAZY} \citep{Brammer2008} template-fitting codes. The catalog reaches a redshift precision of $\sigma_{\text{MAD}} < 0.01(1+z)$ for bright sources ($i < 22.5$\,mag). For the faintest sources ($i < 25$\,mag) the precision is degraded, but still better than  $\sigma_{\text{MAD}} < 0.025(1+z)$. 

\paragraph{IRAC:} To incorporate deep near-infrared data, we use the S-COSMOS catalog \citep{Sanders+07}. This Spitzer Space Observatory Legacy program provides a uniform survey of the entire $2~\text{deg}^2$ area. The catalog includes observations from all seven Spitzer bands: four channels from the Infrared Array Camera (IRAC) at $3.6$, $4.5$, $5.8$, and $8.0~\mu\text{m}$, and three channels from the Multiband Imaging Photometer for Spitzer (MIPS) at $24$, $70$, and $160~\mu\text{m}$. The depth of the survey is characterized by $5\sigma$ sensitivities of $0.9~\mu\text{Jy}$ and $1.7~\mu\text{Jy}$ for the IRAC $3.6$ and $4.5~\mu\text{m}$ channels, respectively. 

\subsubsection{Mid- and far-infrared ancillary data}

\paragraph{MIPS (24~$\mu$m):}  
For mid-infrared observations, we use the COSMOS MIPS $24~\mu\text{m}$ catalog \citep{LeFloch+09}, obtained with an angular resolution of $6\arcsec$, covering an effective area of $1.68~\text{deg}^2$.
This catalog contains $39,413$ sources and represents one of the most complete infrared surveys of the COSMOS field, achieving $90\%$ completeness at a flux density of $\sim 80~\mu\text{Jy}$ and a $1\sigma$ sensitivity of $18~\mu\text{Jy}$. 

\paragraph{\textit{Herschel}/PACS:} 
We used the \textit{Herschel}/PACS data at $100\,\mu$m and $160\,\mu$m from the PACS Evolutionary Probe survey \citep[PEP;][]{Lutz2011}, with angular resolutions of approximately 7.4\arcsec\ and 11.3\arcsec, respectively. The PEP observations cover the full $\sim$2\,deg$^{2}$ COSMOS field. Blind source catalogs were generated using PSF-fitting extraction. The catalogs contain $3,398$ and $2,581$ sources detected at $\mathrm{S/N}\geq5$ at $100\,\mu$m and $160\,\mu$m, respectively. The corresponding sensitivities reach $1\sigma=1.50$\,mJy at $100\,\mu$m and $1\sigma=3.27$\,mJy at $160\,\mu$m.
Note that the quoted sensitivities do not take into account confusion noise.

\paragraph{\textit{Herschel}/SPIRE:} We used the \textit{Herschel} Extragalactic Legacy Project \citep[HELP\footnote{\url{https://hedam.lam.fr/HELP/}.};][]{Shirley+21} SPIRE blind catalogs, which cover multiple extragalactic fields for an effective area of $1270~\text{deg}^2$. The SPIRE blind catalogs in COSMOS contain $12,603$ sources detected independently at $250~\mu\text{m}$, $350~\mu\text{m}$, and $500~\mu\text{m}$ over an area of 5.1\,deg$^2$. The HELP dataset builds from the original \textit{Herschel} Multi-tiered Extragalactic Survey  \citep[HerMES;][]{Oliver+12}, which reaches a $3\sigma$ depth of 8.1\,mJy, 10.7\,mJy and 15.4\,mJy at $250~\mu\text{m}$, $350~\mu\text{m}$, and $500~\mu\text{m}$, respectively in COSMOS. The SPIRE bands have angular resolutions of approximately 18\arcsec, 25\arcsec, and 36\arcsec, respectively.

\paragraph{SCUBA2:} For the submillimeter observations, we used both the SCUBA2-COSMOS survey (S2COSMOS; \citealt{Simpson+19}), taken as part of the East Asian Observatory (EAO) Large Program series, and the SCUBA2 Ultra Deep Imaging EAO Survey (STUDIES; \citealt{Gao2024}). The S2COSMOS catalog \citep{Simpson+19} covers a total area of $2.6~\text{deg}^2$, where the  \textsc{main} (HST/ACS footprint) region achieves a median noise level of $\sigma_{850~\mu\text{m}} = 1.2~\text{mJy/beam}$ over $1.6~\text{deg}^2$, and a $1.0~\text{deg}^2$ supplementary (\textsc{supp}) region with $\sigma_{850~\mu\text{m}} = 1.7~\text{mJy/beam}$, identifying a total of $1,147$ sources at $>4\sigma$ threshold. In contrast, the STUDIES program \citep{Gao2024} targets a smaller area of $450~\text{arcmin}^2$ but reaches deeper sensitivities. The $450~\mu\text{m}$ catalog reaches instrumental noise level of $\sigma_{450~\mu\text{m}} = 0.59~\text{mJy/beam}$, while the $850~\mu\text{m}$ catalog reaches a sensitivity of $\sigma_{850~\mu\text{m}} = 0.09~\text{mJy/beam}$. Above the $\geq 3.5\sigma$ threshold, the catalogs include 479 and 324 sources at $450~\mu\text{m}$ and $850~\mu\text{m}$, respectively. 
SCUBA-2 provides angular resolutions of 9.8\arcsec\ and 14.6\arcsec\ at 450\,$\mu$m and 850\,$\mu$m, respectively.

\subsubsection{Radio ancillary data}

\paragraph{MIGHTEE:}  We made use of the MIGHTEE survey \citep{MeerKat}, which provides radio continuum observations obtained with the South African MeerKAT telescope at a central frequency of 1284\,MHz (L band), with an angular resolution of 8.6\arcsec.
The catalog covers $1.6~\text{deg}^2$ within the COSMOS field through a single pointing. 
The MIGHTEE COSMOS field reaches a thermal noise level below $2~\mu\text{Jy/beam}$, and identifies $9,896$ sources. 

\paragraph{VLA 1.4\,GHz:} In addition to the 3\,GHz catalog \citep{3GHZVLA}, we also use the VLA-COSMOS Joint Catalog \citep{1.4GHzVLA}, which provides deep $1.4$\,GHz (20\,cm) radio continuum observations over the $2~\text{deg}^2$ field. This catalog integrates the VLA-COSMOS Large and Deep Projects, reaching a $1\sigma$ sensitivity of $\approx 12~\mu\text{Jy beam}^{-1}$. The joint catalog comprises 2,865 radio sources detected at a significance of $\geq 5\sigma$, with resolutions of $2.5\arcsec$ for the Deep mosaic and $1.5\arcsec$ for the Large Project.

\subsubsection{The A3COSMOS photometric catalog}
As described in Sect.~\ref{sec: proxy_identification}, the A3COSMOS survey involved two catalogs (blind and prior extraction) built from all available public ALMA observations at (sub)millimeter wavelengths. In both catalogs, one source might have multiple observations in different programs and/or wavelengths. Therefore, we built a photometric A3COSMOS master catalog (available in the N2CLS database) for the N2CLS sources with counterpart in any of these catalogs by collecting the different observations for the unique sources. This photometric catalog is based on the two master catalogs constructed for A3COSMOS, corresponding to the different flux-extraction methods (see Sect.~\ref{sec: proxy_identification}).

\subsection{Multiwavelength SED construction}\label{sect: SED}

We built the multiwavelength SED of each N2CLS source by adding the photometry of each ancillary catalog. For COSMOSWeb and COSMOS2020 (Classic), we exclude the narrow bands. To ensure a robust SED reconstruction, we carried out a visual inspection of each source using the multiwavelength cutouts (which are available in the N2CLS database), where we collected the available public mosaics of each of the ancillary catalogs. For SCUBA2 at $850~\mu\text{m}$, we used the mosaic from the SCUBA2 Cosmology Legacy Survey (S2CLS) presented in \cite{Geach2017}, which covers an area of 2.2\,deg$^2$ in 416\,h with a median noise level of $\sigma_{850~\mu\text{m}} = 1.6 ~\text{mJy/beam}$. The $850~\mu\text{m}$ map presented in \cite{Simpson+19}, which added 223\,h of observations to S2CLS, is not publicly available.

Through visual inspection of the cutouts, we clean the SEDs of our sources by removing photometric measurements that may be contaminated by blended sources, particularly in bands with coarse spatial resolution, such as SPIRE, whose PSF (18–36\,arcsec) is several times larger than that of optical and near-infrared observations. 
This is especially critical for N2CLS sources with multi-component associations (41/323), found through the proxy identification presented in Sect.~\ref{sec: proxy_identification}. Photometric measurements from catalogs characterized by relatively large beam sizes (namely NIKA2, SCUBA2, PACS, SPIRE, MIPS, and IRAC) were removed when visual inspection of the corresponding cutouts revealed clear blending with nearby sources. For blended sources, we adopted flux measurements from the super-deblended catalog \citep{Jin+18,Liu+18} when available, provided that the deblended flux satisfies ``$\mathrm{flux} > 3 \times \mathrm{flux\ uncertainty}$.''
A few example multiwavelength cutouts of N2CLS sources are shown in Appendix\,\ref{app:cutouts} for illustration, while the full set of cutouts is provided on the data release page.
The SEDs will be analyzed in detail in Berta et al. (in prep), to derive statistical constraints on the physical properties of the N2CLS galaxies.

\section{Redshift distribution \label{sec:z_distrib}}

\subsection{N2CLS galaxies}\label{sec:n2cls_galaxies}
The redshift distribution provides key constraints on galaxy evolution models. At (sub)millimeter wavelengths, the average redshift of detected sources is expected to increase with wavelength due to the negative $k$-correction, although this trend also depends on survey depth (e.g., \citealt{Bethermin+15}). The N2CLS survey thus offers a valuable opportunity to test this behavior and place further constraints on galaxy evolution. Consequently, determining reliable redshifts for our sources is a crucial step of this analysis.

After we identified the accurate position of the N2CLS sources using high-resolution data (see Sect.~\ref{sec: proxy_identification}) and their multiwavelength counterparts (see Sect.~\ref{sec: multi_counterpart}), we gathered all the redshift information available in the literature. Spectroscopic redshifts are compiled from multiple studies (including the recent compilation by \citealt{Khostovan+26}) and are listed in the source catalog presented in the N2CLS database. In addition, we confirmed a spectroscopic redshift for N2CO\_1\_241 through our NOEMA follow-up observations (W23CJ; PI: A. Beelen). We detect an emission line at 134.382\,GHz that we identify as CO(4--3), corresponding to a spectroscopic redshift of $z_{\rm spec}=2.431$. This value is based on the photometric redshift estimates from COSMOS2020 ($z_{\rm phot}$=2.2$\pm$0.28) and COSMOSWeb ($z_{\rm phot}$=2.5$\pm$ 0.02). Photometric redshifts are primarily taken from the COSMOS2020 catalog \citep{COSMOS2020}. When not available in COSMOS2020, we complemented with the COSMOSWeb catalog \citep{shuntov+25}, and with the super-deblended catalog \citep[five galaxies,][]{Jin+18,Liu+18}. This priority order was adopted due to an observed systematic bias in the redshift distribution derived from the COSMOSWeb photometric redshifts of our sources. A detailed discussion is provided in Appendix~\ref{app: photo_redshift}. We also used the NED database\footnote{\url{https://ned.ipac.caltech.edu/}.} to check some individual sources. 

Tables~\ref{tab: counterparts_redshift_95} and \ref{tab: counterparts_redshift_80} present a summary of the redshift type per proxy for the high purity (95\%) and intermediate purity (80\%-95\%) subsamples, respectively. At 95\% purity, there are 6 proxies without any redshift, plus 16 N2CLS sources without proxy identification, which represent $\sim$8\% of the total sample (proxy + unidentified N2CLS sources). Among the identified proxies, 109 ($\sim$40$\%$) have spectroscopic redshifts, and 154 ($\sim$57$\%$) have photometric redshifts. For the intermediate purity sample, just one proxy does not have redshift, 27 ($\sim$40$\%$) have spectroscopic redshifts, and 40 ($\sim$ 59$\%$) have photometric redshifts. We have no redshift for $\sim$22$\%$ of the total intermediate purity sample, 18 N2CLS sources without proxy, and one counterpart. 

\begin{table*}
\centering
\caption{Counterparts per proxy and redshift types for the $\geq$95\% purity sample (S/N$\geq$4.5 at 1.2\,mm and $\geq$4.6 at 2\,mm).}
\label{tab: counterparts_redshift_95}
\resizebox{\linewidth}{!}{
\begin{tabular}{ccccccccccccc}
\hline
\multirow{2}{*}{Redshift Type}
& \multicolumn{2}{c}{AS2COSMOS} 
& \multicolumn{2}{c}{A3COSMOS} 
& \multicolumn{2}{c}{NOEMA} 
& \multicolumn{2}{c}{CHAMPS} 
& \multicolumn{2}{c}{Radio} 
& \multirow{2}{*}{Total}
& \multirow{2}{*}{No proxy}\\
\cline{2-11}
 & Single & Multiple 
 & Single & Multiple 
 & Single & Multiple 
 & Single & Multiple 
 & Single & Multiple 
 &\\
\hline
No-z     &  --  & -- & -- &  2 & 1 & -- & 1  & --  & 2  & -- &   6  & 16 \\
Photo-z  & 15  & 15 & 45 & 16 & 9 & 4  & 24 & 3   & 21 &  2 &  154 & -- \\
Spec-z   & 21  & 21 & 32 &  8 & 3 & -- & 9  & 3   & 9  &  3 &   109 & -- \\
\hline
Total    & 36   & 36 & 77 & 26 & 13 & 4 & 34 & 6   & 32 & 5  & 269 & 16 \\
\hline
\end{tabular}
}
\end{table*}

\begin{table*}
\centering
\caption{Same as Table~\ref{tab: counterparts_redshift_95} but for purity between 80\% and 95\% (i.e., 3.9$<$S/N$<$4.5 at 1.2\,mm and 3.9$<$S/N$<$4.6 at 2\,mm).}
\label{tab: counterparts_redshift_80}
\resizebox{\linewidth}{!}{
\begin{tabular}{ccccccccccccc}
\hline
\multirow{2}{*}{Redshift Type}
& \multicolumn{2}{c}{AS2COSMOS} 
& \multicolumn{2}{c}{A3COSMOS} 
& \multicolumn{2}{c}{NOEMA} 
& \multicolumn{2}{c}{CHAMPS} 
& \multicolumn{2}{c}{Radio} 
& \multirow{2}{*}{Total}
& \multirow{2}{*}{No proxy}\\
\cline{2-11}
 & Single & Multiple 
 & Single & Multiple 
 & Single & Multiple 
 & Single & Multiple 
 & Single & Multiple 
 &\\
\hline
No-z     &  --  & -- & -- & -- & -- & -- & -- & --   & 1 & -- &   1 & 18 \\
Photo-z  & 1    & -- & 13 & 6  & -- & -- & 10 & --  & 9  &  1 &   40 & -- \\
Spec-z   & --   & -- & 11 & 4  & 1  & -- & 4  & --   & 6  &  1 &   27 & -- \\
\hline
Total    & 1    & -- & 24 & 10 & 1  & -- & 14 & --  & 16 &  2  &  68  & 18 \\
\hline
\end{tabular}
}
\end{table*}

\begin{figure}[!t]
\centering
\includegraphics[width=1\columnwidth]{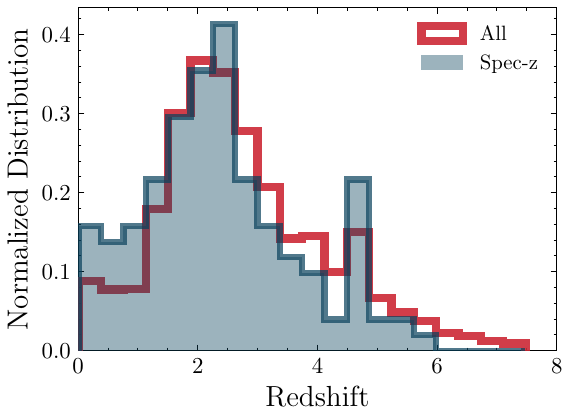}
\caption{Normalized redshift distribution of the N2CLS sample, accounting for redshift uncertainties through a Monte Carlo approach (see Sect.~\ref{sec:n2cls_galaxies}). The red histogram shows the distribution of all redshifts, while the blue histogram corresponds to the spectroscopically confirmed subsample.}
\label{fig: redshift_dist}
\end{figure}

In Fig.~\ref{fig: redshift_dist}, we show the redshift distribution of the N2CLS galaxies. The median redshift of the full sample is z=2.51$_{-0.01}^{+0.01}$, with a median absolute deviation (MAD) of $0.825_{-0.3}^{+0.3}$, where the uncertainties are estimated using a Monte Carlo approach in which each redshift is modeled as a Gaussian probability density function (PDF) centered on the measured value, with $\sigma$ as the reported uncertainty. We identify 19 galaxies at z$>$4, including four at z$>$5, with spectroscopic redshift, all of them have a strong detection at 1.2\,mm (S/N$>$5.5), although three lack detection at 2\,mm (N2CO\_1\_96, N2CO\_1\_148, N2CO\_1\_158\_a). The highest spectroscopic redshift in our sample is N2CO\_1\_49 (A3COSMOS proxy) at z=5.85 \citep{Jin+19}, with corrected flux density of $\rm S_{corr}$=3.01$_{-0.50}^{+0.54}$\,mJy (S/N=9) at 1.2\,mm, and $\rm S_{corr}$=0.63$_{-0.15}^{+0.18}$\,mJy (S/N=6.7) at 2\,mm. This source is undetected at short wavelengths and is detected only in JWST/NIRCam F277W and F444W, as well as in MIRI F770W.

\begin{figure*}[h]
\centering
\includegraphics[height=0.22\textwidth]{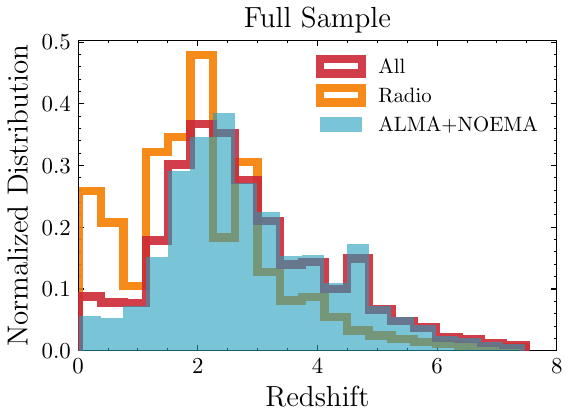}
\includegraphics[height=0.22\textwidth]{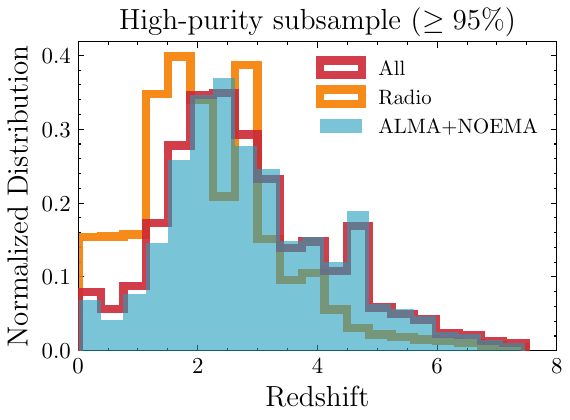}
\includegraphics[height=0.22\textwidth]{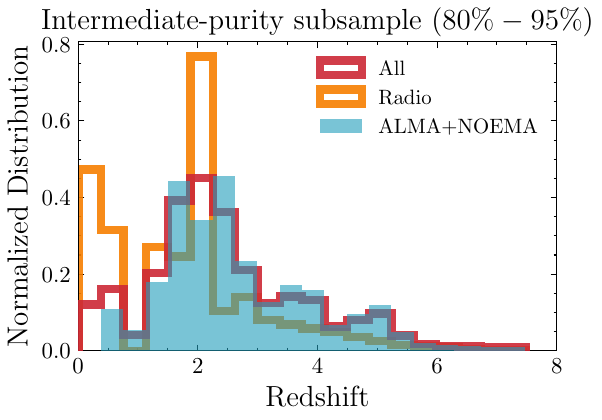}
\caption{Normalized redshift distributions of the N2CLS sample for each proxy type, obtained from either (sub)millimeter or radio interferometry, with redshift uncertainties accounted for through a Monte Carlo approach (see Sect.~\ref{sec:n2cls_galaxies}). \textit{From left to right}: Full sample, high-purity subsample ($\geq95\%$), and intermediate-purity subsample (80–95\%). The numbers of objects in the full sample, the high-purity subsample, and the intermediate-purity subsample respectively are 330, 263, and 67. Of these, 278, 228, and 50 have been identified through (sub)millimeter interferometry, while 52, 35, and 17 have been identified through radio interferometry, respectively.
}
\label{fig: redshift_dist_proxy}
\end{figure*}

\begin{figure}[!t]
\centering
\includegraphics[width=1\linewidth]{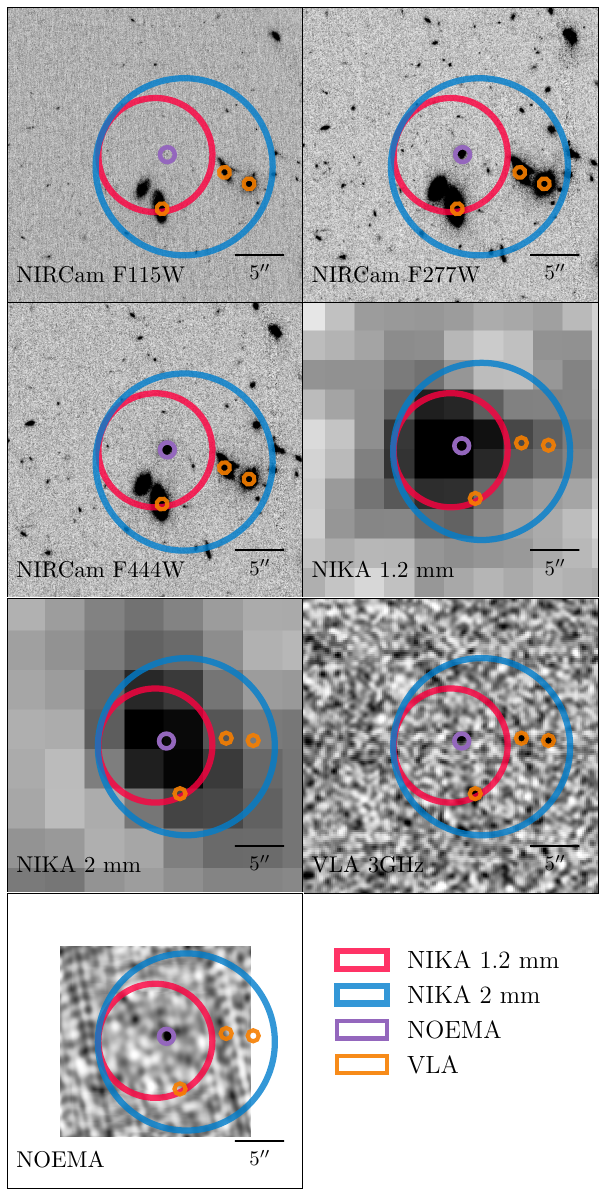}
\caption{Cutouts of source N2CO\_1\_99 illustrating a misidentification based on a radio counterpart. The NOEMA 241\,GHz continuum source, highlighted in purple, reveals the right counterpart at z$_{\rm spec}$=4.916. In contrast, the VLA 3\,GHz radio source (orange) located 5.5\arcsec\ from the NIKA2 1.2\,mm position and lying within its FWHM corresponds to a late-type galaxy at z$_{\rm spec}$=0.71, which would lead to an incorrect association if used as the counterpart. Two additional VLA 3\,GHz sources are also nearby, but they lie outside the NIKA2 1.2\,mm beam. The NIKA2 1.2\,mm and 2\,mm beams are shown in red and blue, respectively.}
\label{fig: noema_vla_iden}
\end{figure}

Figure~\ref{fig: redshift_dist_proxy} compares the redshift distributions by proxy type, grouping proxies based on (sub)millimeter interferometry (AS2COSMOS, A3COSMOS, NOEMA, CHAMPS) and radio-based identifications (VLA 3\,GHz, VLA 1.4\,GHz, MeerKAT) for the full sample and for the different purity ranges. In all cases, the redshift distributions derived from radio-identified counterparts exhibit lower median values (z$\sim$1.9–2.0) compared to those based on (sub)millimeter interferometric identifications (z$\sim$ 2.4-2.8). This behavior is consistent with previous studies showing that identifications relying on radio tend to bias samples toward lower redshifts \citep[e.g.,][]{Chapman+05, Smolcic2012, Alberts2013, Brisbin+17}.
Source N2CO\_1\_99, shown on Fig.~\ref{fig: noema_vla_iden}, is a clear example of a misidentification based on a radio proxy: within the N2CLS 1.2\,mm beam, the VLA 3\,GHz source J100118.74+022347.8 (orange beam) is located 5.5\arcsec\ away from the center position of the N2CLS source and corresponds to a late-type galaxy at z$_{\rm spec}$=0.71 \citep{Hasinger+18,Khostovan+26}. However, high-resolution N2CLS NOEMA follow-up observations identify the correct counterpart (purple beam) at z$_{\rm spec}$=4.916 \citep{Forrest+25,Khostovan+26}, located $\sim$1\arcsec\ from the N2CLS 1.2\,mm position. Therefore, radio-based identifications can lead to incorrect low-redshift associations for high-redshift dusty galaxies, and therefore such identifications should be treated with caution in the absence of (sub)millimeter interferometric observations. In the high-purity subsample (>95\%), $\sim$21$\%$ (50/243) lack (sub)millimeter interferometric positional proxies; of these, 34 are identified using radio proxies.

\subsection{Comparison with models}

As discussed in Sect.~\ref{sect:cat_char} (see also \citetalias{Bing+23}) and demonstrated by \citet{Bethermin+26} \citepalias[hereafter][]{Bethermin+26}, the SIDES simulations play a central role in this work, both for correcting the measured source fluxes and for interpreting the observed results.
When comparing the N2CLS observations to models, and because the N2CLS observations have an angular resolution of $\sim$11\arcsec, one has to consider the ``blob'' catalogs (and not the galaxy catalogs; see Sect.~\ref{sect:cat_char}), which are the only catalogs in where a flux cut can be defined properly for a given chosen purity (or completeness).  Such a blob catalog is available only for the SIDES simulations, for which we generated noiseless maps, convolved with the NIKA2 beam, and subsequently extracted sources to produce the corresponding blob catalog, thereby accounting for source blending effects. We also use the SIDES galaxy catalog \citep{Bethermin+17, Gkogkou+23}, which contains all galaxies in the simulation. From this catalog, we select sources according to the flux limits of our observations.

We first compared our redshift distribution with SIDES (both ``blob'' and galaxy catalogs) and then extended to other models in the literature. 
To compare with models, we only considered the N2CLS sources at 1.2\,mm. We assumed a flux cut of S/N$\times \sigma$ adopting S/N=4.5 ($\geq 95\%$ high purity subsample) and $\sigma$=315\,$\mathrm \mu$Jy (see Sect.~\ref{subsec: data_redu}). 
The redshift distribution of the 2\,mm sources is described in \citetalias{Bethermin+26}. We update this analysis with improved redshift completeness, with results presented in Appendix\,\ref{App:2mm_z}.

\subsubsection{Redshift distribution in SIDES}\label{sec: z_distribution_sides}

\begin{figure}[!t]
\centering
\includegraphics[width=0.47\textwidth]{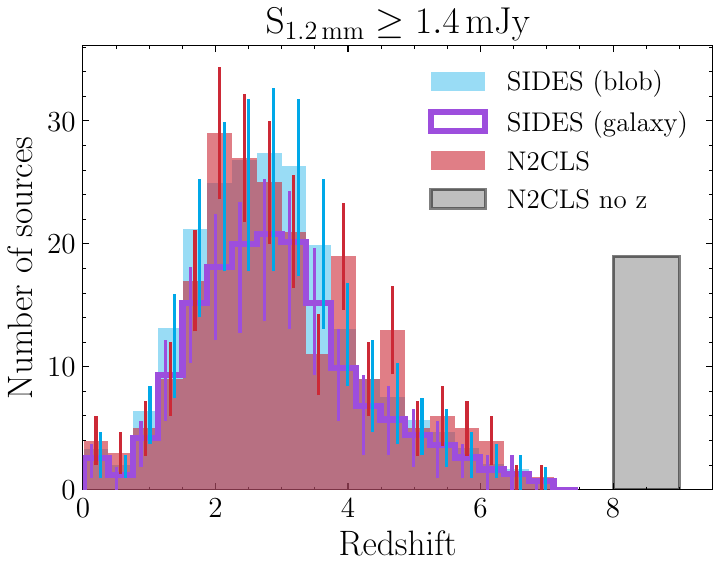}
\includegraphics[width=0.47\textwidth]{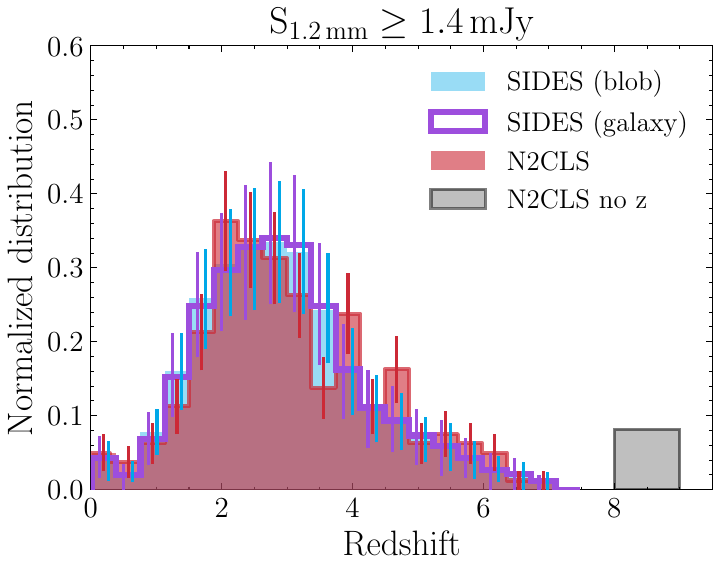}
\caption{Redshift distribution of the N2CLS high-purity  ($\geq95\%$, S/N$>4.5$) 1.2\,mm subsample (red histograms). \textit{Top}: Number counts. \textit{Bottom}: Normalized distribution (normalized to unit area, $\int n(z)\, dz = 1$). When there are multiple proxies within the same beam, the redshift of the brightest component is used. The gray histogram at z=8-9 shows the fraction ($\sim$8$\%$) without identification (14 N2CLS sources) and without redshift (five proxies: one from A3COSMOS, one from CHAMPS, one from NOEMA, and two from VLA 3\,GHz).
The uncertainties of N2CLS correspond to a Poisson law. The purple and cyan histograms were derived from the SIDES galaxy catalog \citep{Gkogkou+23} and the SIDES blob catalog, respectively, after applying the flux cut $\rm S_{1.2\,mm} \geq 1.4$\,mJy, which corresponds to the flux limit of the N2CLS high-purity subsample. The uncertainties in both SIDES catalogs reflect the field-to-field variance measured over $33.6\times33.6$\,arcmin$^2$ subfields.}
\label{fig:redshift_dist_sides}
\end{figure}

In Fig.~\ref{fig:redshift_dist_sides}, we compare the redshift distribution of the 1.2\,mm sources in the high-purity subsample ($\geq 95\%$ purity) with the SIDES blob (see Sect.~\ref{sect:cat_char}) and galaxy catalogs \citep{Gkogkou+23}. For multi-counterpart sources, we assign the redshift of the brightest component. This applies to 35 N2CLS sources that are resolved into 75 counterparts. The final subsample comprises 214 sources with assigned redshifts (180 single-component and 34 multi-component systems), and 19 sources (14 unidentified) without redshift information. We exclude the source N2CO\_1\_155, which is resolved into two A3COSMOS counterparts, because the brightest component lacks a redshift measurement. To compare the N2CLS observations with the SIDES predictions, we apply the same flux cut to the SIDES catalogs ($\rm S_{1.2\,mm} \geq S/N \times \sigma_{\rm noise~1.2\,mm} = 1.4$\,mJy). Although this assumes a uniform survey depth, the redshift distribution is only weakly sensitive to the exact flux limit \citep{Bethermin+15}.  We account for resolution effects using the SIDES blob catalog, in which multiple galaxies may fall within a single NIKA2 beam; in such cases, we assign the redshift of the brightest 1.2\,mm galaxy
\citep{Bethermin+17, Bethermin+26}. We computed the field-to-field variance by dividing both SIDES catalogs into $33.6\times33.6$\,arcmin$^2$ subfields (330 in total).

The normalized redshift distributions of the SIDES galaxy and blob catalogs are very similar, which shows that the distribution is not sensitive to the resolutions, unlike the number counts, as also shown for 2\,mm-selected sources \citepalias[e.g.,][]{Bing+23, Bethermin+26}. The N2CLS high-purity subsample exhibits a similarly shaped redshift distribution, with a small excess at 3.5$<$z$<$5 that remains within the $1\sigma$ uncertainty.
We measure a median redshift of $2.8 \pm 0.1$, with the uncertainty estimated from Poisson law. In comparison, SIDES predicts median values of $2.8 \pm 0.1$ for both the blob and galaxy catalogs.

We performed two-sample Kolmogorov–Smirnov (KS) tests to assess whether the observed N2CLS redshift distribution is statistically consistent with the SIDES predictions, as well as to directly compare the two SIDES catalogs. The KS test measures the maximum absolute difference ($D$) between the cumulative distribution functions (CDFs) of two samples, with the associated $p$-value quantifying the probability of obtaining a $D$ value at least as large as the observed one under the null hypothesis that both samples are drawn from the same parent distribution. Typically, a $p$-value $>5\%$ is taken as insufficient evidence to reject the null hypothesis, i.e., that the two samples are consistent with being drawn from the same parent distribution \citep{WallJenkins2012}.
When comparing the SIDES distributions with N2CLS, no statistically significant difference is found, with $p$-values of 0.17 and 0.24 for the blob and galaxy catalogs, respectively. In both cases, the maximum deviation ($D \sim 0.078$) is at z$\simeq$3.7, where the cumulative N2CLS distribution lies slightly below SIDES. However, the amplitude of this difference is within the range expected from statistical fluctuations.

Since $\sim$60$\%$ of our sources rely on photometric redshifts, we verified 
that these results are not driven by the redshift uncertainties. Following 
the same Monte Carlo scheme adopted in Sect.~\ref{sec:n2cls_galaxies}, we 
generated $10^3$ realizations of the N2CLS redshift distribution, drawing 
each redshift from a Gaussian PDF centered on the measured value with a 
$\sigma$ given by its reported uncertainty. Repeating the KS test on each 
realization yields $D = 0.069^{+0.013}_{-0.012}$ and 
$p = 0.24^{+0.22}_{-0.14}$ for the blob catalog, and 
$D = 0.067^{+0.012}_{-0.009}$ and $p = 0.27^{+0.18}_{-0.15}$ for the galaxy 
catalog, where the uncertainties are taken from the 16th--84th percentiles 
of the realizations. These values are consistent with those obtained from 
the unperturbed redshifts, and the null hypothesis is rejected ($p$-value$<$0.05) in only $4\%$ and $2\%$ of the realizations, respectively. Accounting 
for the photometric redshift uncertainties therefore does not change our 
conclusions.

The KS test analysis therefore shows that the N2CLS redshift distributions at 1.2\,mm (and 2\,mm as shown in Appendix\,\ref{App:2mm_z}) are statistically consistent with the SIDES predictions \citepalias[see also][]{Bethermin+26}. This confirms that SIDES provides a reliable reference model for the redshift distributions of bright 1.2\,mm- and 2\,mm-selected sources.

\subsubsection{Redshift distribution in other models \label{sect:z_model}}

We gathered predictions from galaxy evolution models that provide suitable and publicly available (sub)millimeter galaxy catalogs for comparison with our observations at 1.2\,mm. These predictions are presented in the following list:

\begin{itemize}
\item SHARK, which is a semi-analytical model introduced in \cite{Lagos2018}. This model applies semi-analytical recipes to describe the evolution of galaxies in dark-matter halos from numerical simulations. The dust emission and (sub)millimeter fluxes of galaxies are predicted based on their gas content and metallicity using a framework described in \cite{Lagos2019, Lagos2020}. 

\item The semi-empirical model of \cite{Popping+20}, which assigns star formation rates in dark matter halos following the SFR-halo relation from the UNIVERSEMACHINE \citep{Behroozi2019}, and then uses empirical relations to estimate the dust mass and obscured star formation.
The 850\,$\mu$m and 1.1\,mm flux density of galaxies are derived following the fits presented in \cite{Hayward2011} and \cite{Hayward2013}, as a function of galaxy SFR and dust mass.

\item The backward evolution model of \cite{Zavala2021}, which combines a parameterized evolving galaxy IR luminosity function with the thermal SED properties of galaxies’ dust emission to make predictions for galaxy (sub)millimeter surveys.

\item The modeled submillimeter galaxies in the FLAMINGO (Full-hydro Large-scale structure simulations with All-sky Mapping for the Interpretation of Next Generation Observations) large-volume hydrodynamical cosmological simulations \citep{Kumar+25}. For this model, we do not have galaxy catalogs but a histogram of the redshift distribution for $S_{850\,\mu m}>3$\,mJy, that we rebinned to the same bin ranges as those used for the other models.
\end{itemize}

\begin{figure*}[!t]
\centering
\includegraphics[width=0.24\textwidth]{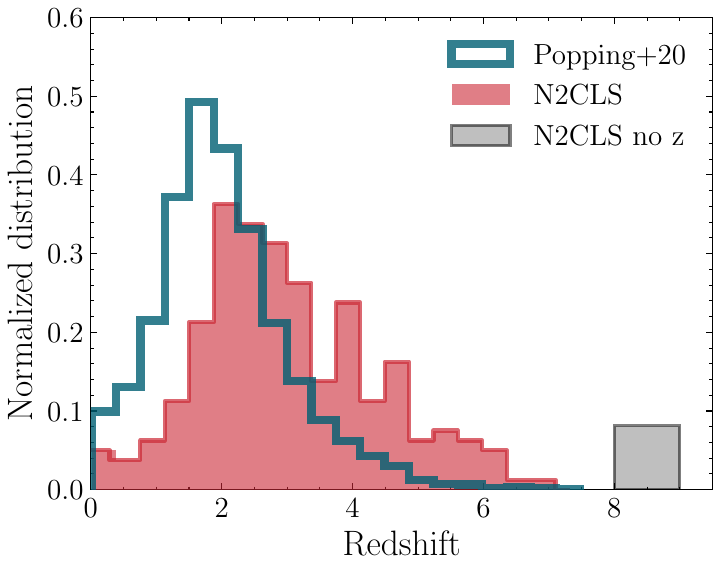}
\includegraphics[width=0.24\textwidth]{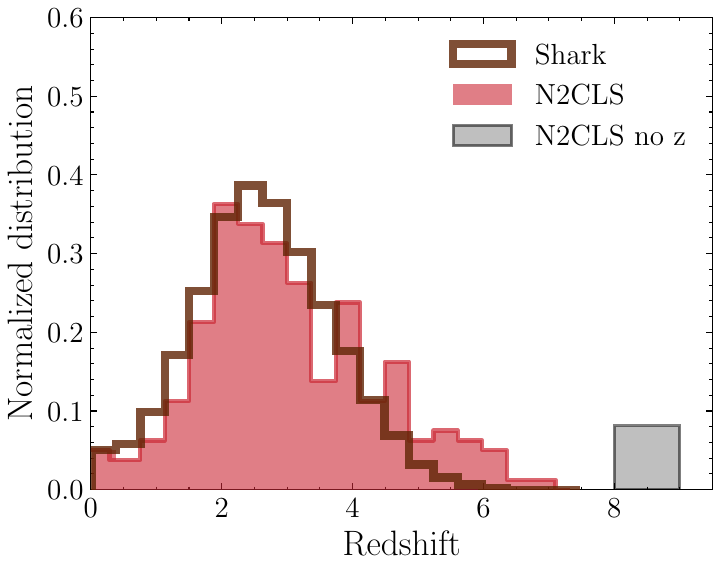}
\includegraphics[width=0.24\textwidth]{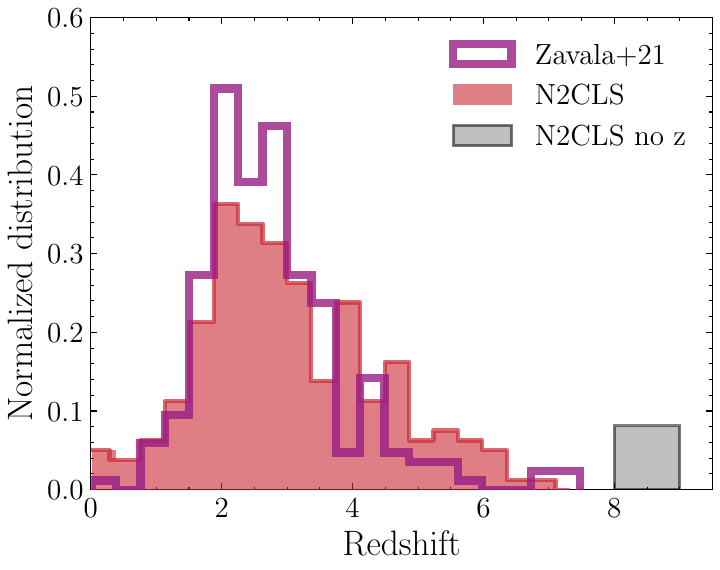}
\includegraphics[width=0.24\textwidth]{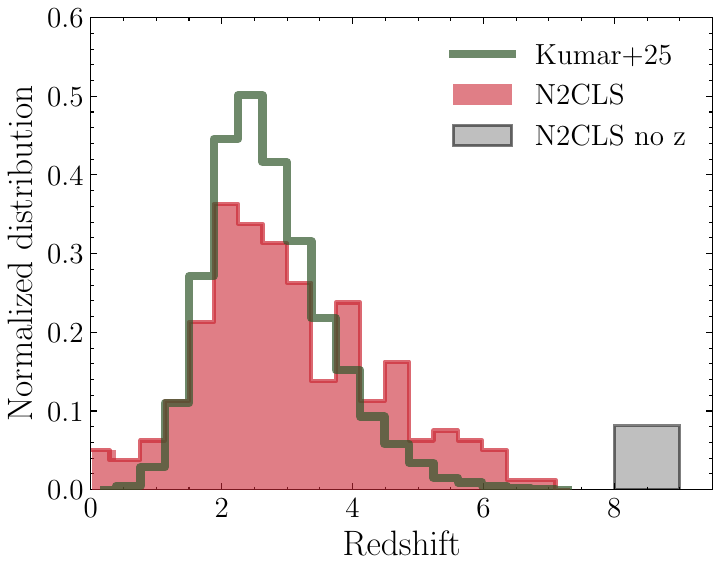}
\caption{Normalized redshift distributions of the N2CLS high-purity subsample at 1.2\,mm compared with predictions from different galaxy evolution models. From left to right: \citet{Popping+20}, \textsc{SHARK} \citep{Lagos2018}, \citet{Zavala2021}, and \citet{Kumar+25}. The distributions from the different models were derived after applying the flux cut $\rm S_{1.2\,mm} \geq 1.4$\,mJy (except for \cite{Kumar+25}, where their 850\,$\mu$m flux cut corresponds to 
$S_{1.2\,mm} \geq 1.20-1.71$ over $1\le z \le 6$; see Sect.\,\ref{sect:z_model}), which corresponds to the flux limit of the N2CLS high-purity subsample. All distributions have been normalized to unit area ($\int n(z)\, dz = 1$).}
\label{fig: redshift_dist_models}
\end{figure*}

For the models of \cite{Popping+20} and \cite{Kumar+25}, we had to convert the 1.2\,mm flux cut, $\rm S_{1.2\,mm} \geq 1.4$\,mJy, to 1.1\,mm and 850\,$\mu$m, respectively. For that purpose, we use the SIDES galaxy catalog. The ratio $S_{1.2\,mm}/S_{1.1\,mm}$ shows only a weak variation, from 0.80 to 0.84 over $1\le z \le6$, and can thus be considered approximately constant.
On the contrary, the ratio $S_{1.2\,mm}/S_{850\,\mu m}$ varies from 0.40 to 0.57 over $1\le z \le6$, corresponding to an effective flux cut at 850\,$\mu$m between 3.54 and 2.49\,mJy, respectively. This range brackets the 3\,mJy flux cut used for the redshift distribution from \cite{Kumar+25}. However, this variation may affect the shape of the redshift distribution, and the comparison should therefore be treated with caution.

Moreover the comparison with the models is performed using their individual-galaxy catalogs (rather than blob catalogs), but we do not expect the treatment of multiplicity to significantly affect our results. Indeed, the normalized redshift distributions of the SIDES galaxy and blob catalogs are very similar (Sect.\,\ref{sec: z_distribution_sides}). Moreover, in our observed sample, at most ten sources (less than 5\% of the full sample) could be significantly affected by assigning the redshift of the brightest counterpart to an N2CLS source: these correspond to multiple identifications with different redshifts for which the 1.2\,mm flux ratio between the brightest and secondary counterpart is either below two (four sources) or cannot be estimated (ten sources).

In Fig.~\ref{fig: redshift_dist_models}, we compare the normalized N2CLS redshift distribution of the high-purity subsample with the models we presented above, after applying the same flux cut we used for N2CLS and SIDES sources. We measure a median redshift of $z = 1.9$ for the model of \citet{Popping+20}, significantly lower than N2CLS (z=2.8$\pm$0.1). In contrast, both SHARK and the backward evolution model of \citet{Zavala2021} predict median redshifts of z$\simeq$2.6, in closer agreement with our measurements. For the FLAMINGO model \citep{Kumar+25}, individual galaxy redshifts are not available. We therefore estimated the median directly from the redshift histogram, obtaining z$\sim$2.6. In addition, we reconstruct the CDF from the histogram and draw random realizations via inverse sampling, which yields a consistent median redshift of z=2.5.

The comparison of the N2CLS redshift distribution with \citet{Popping+20}, using the KS test, yields a very large maximum absolute difference ($D \simeq 0.37$) and an extremely low $p$-value ($p\sim 10^{-25}$), which indicate that the two distributions are statistically different. This is also clear in the substantially lower median redshift of the model compared to N2CLS, as well as differences in the overall shape of the distribution. For SHARK, the KS test results in $D \simeq 0.14$ and $p \simeq 4 \times 10^{-4}$. Although the SHARK median redshift (z$\simeq$2.6) is closer to our observed value, the low $p$-value indicates that the N2CLS and SHARK distributions are not drawn from the same parent distribution. The largest deviation is at z$\simeq$3.7–3.8, where the cumulative N2CLS distribution lies below the model prediction. A similar conclusion is reached for the backward evolution model of \citet{Zavala2021}, for which we find $D \simeq 0.17$ and $p \simeq 3.5 \times 10^{-3}$. Despite its median redshift matching that of N2CLS, the test indicates a statistically significant difference with respect to the N2CLS distribution. For the \citet{Kumar+25} model, we perform a Monte Carlo analysis by generating 1000 random realizations of the redshift distribution via inverse sampling of the histogram and apply the test. The resulting KS statistics have a mean value of $D \simeq 0.17$, with a mean $p$-value of $\sim$6$\times 10^{-5}$. These results indicate that, despite a median redshift (z$\simeq$2.5) broadly consistent with the observations, the shape of the redshift distribution does not match with N2CLS.

The KS test analysis shows significant differences in the redshift distributions of the models with N2CLS for the 1.2\,mm high purity subsample, especially for \citet{Popping+20}. Despite the predicted median redshift z$\sim$2.5-2.6 of SHARK (\citealt{Lagos2018}), \citet{Zavala2021}, and \citet{Kumar+25} being consistent with N2CLS ($z=2.8\pm0.1$), none of the models reproduce the CDF of our sample, with a maximum absolute deviation around z$\sim$3.6-3.8 \citep[z$\sim$2.4 for ][]{Popping+20}.

\section{Unidentified and 850\,$\rm \mu$m-dark galaxy populations \label{sec:discussion}}
A full analysis of the N2CLS source SEDs will be presented in Berta et al. (in prep.). The comparison between N2CLS and Ex-MORA \citep{Long2026} sources at 2\,mm is discussed in \citet{Bethermin+26}. Here, particular attention is paid to sources without secure counterparts at other wavelengths, including potential “SCUBA2-dark” objects.

\subsection{Unidentified N2CLS galaxies \label{sect:unidentified}}

Despite the rich multiwavelength datasets available in the COSMOS field and the existence of high-resolution interferometric observations, a fraction of the N2CLS sources remains without a robust identification. Out of the 323 sources in the master N2CO catalog, 34 ($\sim$11\%) lack an identified proxy: 3 are detected in both bands, 25 only at 1.2\,mm, and 6 only at 2\,mm (see Sect.~\ref{sec: proxy_identification}). This corresponds to a high success rate of $\sim$89\% in our full sample. However, the follow-up strategy is biased toward the brightest detections in order to secure interferometric observations, which likely leads to incompleteness for faint, low-S/N sources that could potentially lie at high redshift.

In our sample, all unidentified sources have $\rm S/N_{1.2\,mm}\lesssim6$ and $\rm S/N_{2\,mm}\lesssim5.2$. Therefore, the fraction of unidentified sources is lower in the high-purity subsample ($16/243\sim7\%$; see Table~\ref{tab: identification_95_purity}) than in the intermediate-purity subsample ($18/80\sim22\%$; see Table~\ref{tab: identification_80_purity}). 
For the high-purity subsample, the fraction of unidentified sources at 1.2\,mm is $11/233\sim5\%$ when considering only those without a 2\,mm counterpart, in agreement with the expected spurious fraction at this purity level. Indeed, in total, there are 14 unidentified sources at 1.2\,mm, but 3 of them are also detected at 2\,mm, making them less likely to be spurious. At 2\,mm, the unidentified fraction is lower than expected ($3/90\sim3\%$), and one of these sources is also detected at 1.2\,mm. At intermediate purity, the fraction of unidentified sources increases to $14/68\sim20\%$ at 1.2\,mm, where all unidentified sources are solely detected in this band. At 2\,mm, the unidentified fraction is $6/34\sim18\%$, although 2 of these sources have counterparts at 1.2\,mm. At this purity level, the incompleteness in the proxy-identification process contributes to the higher fraction of unidentified sources. 

\subsection{SCUBA2 850\,$\mu$m dark galaxies}
The N2CLS COSMOS field is entirely covered by SCUBA2 850\,$\rm \mu$m observations from S2COSMOS \citep{Simpson+19} and STUDIES \citep{Gao2024}.
However, a large number of N2CLS sources do not have a SCUBA2 counterpart (see Tables~\ref{tab: numbers_high} and \ref{tab: numbers_low}): 79 sources (with either a single proxy for identification or no identification) plus three N2CLS sources with two identifications, each from A3COSMOS (N2CO\_1\_141, N2CO\_1\_237, and N2CO\_1\_295). Figure~\ref{fig:N2CLS_no850} shows the locations of these sources on the 850\,$\rm \mu$m instrumental noise map from S2COSMOS. Most of them lie outside the very deep SCUBA2 central region.

The redshift distribution of all N2CLS sources (including multiple identifications) was compared with that of sources with a single counterpart but no SCUBA2 850\,$\rm \mu$m detection. No clear trend is observed, even when restricting the sample to high-purity sources and/or to 1.2\,mm sources.

\begin{figure}
    \centering
    \includegraphics[width=0.46\textwidth]{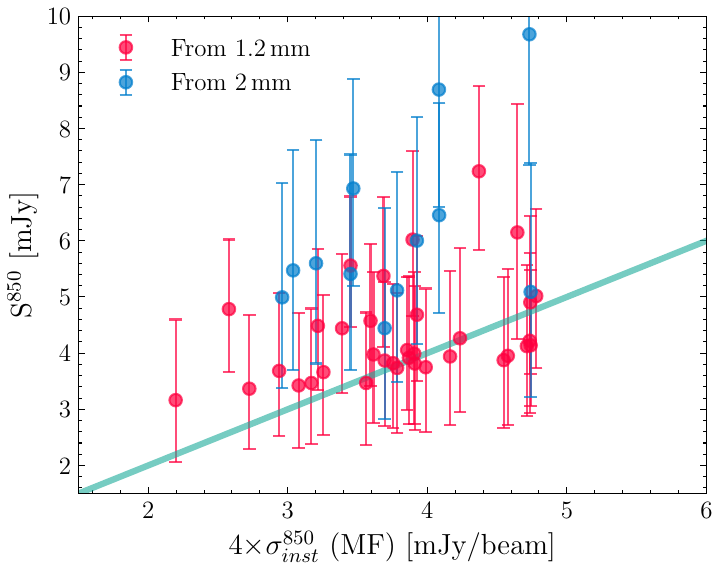}
    \includegraphics[width=0.46\textwidth]{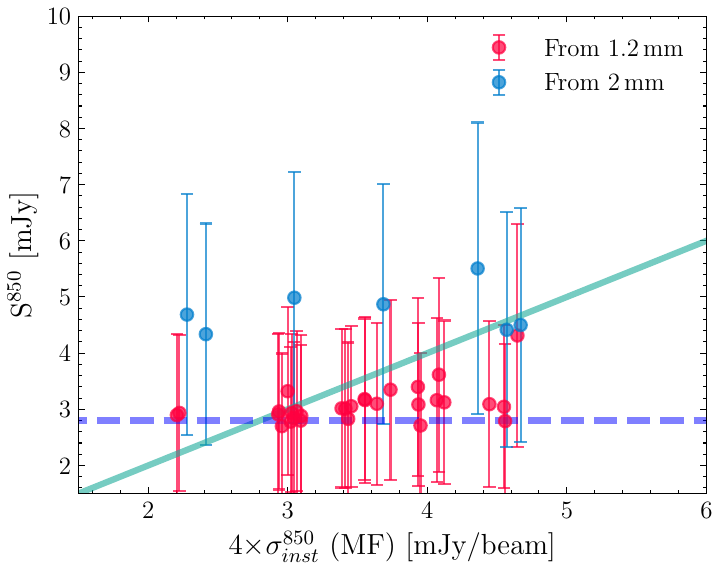}
    \caption{Expected 850\,$\mu$m fluxes computed from the 1.2 and/or 2\,mm N2CLS deboosted fluxes and compared with the 4$\sigma$ threshold applied for the detection of 850\,$\mu$m  sources. All sources above the green line should have been detected by S2COSMOS. \textit{Top}: N2CLS sources with a purity $>$95\%. \textit{Bottom}: N2CLS sources with a purity between 80\% and 95\%. The dashed blue line corresponds to S/N=4 in the deepest S2COSMOS region ($\sigma^{850}_{\rm inst}$=0.7\,mJy/beam). At such a depth, all N2CLS sources should have been detected (i.e., they lie above the line).}
    \label{fig:Expect_S850}
\end{figure}

As in Sect.~\ref{sect:z_model}, we use the SIDES galaxy catalog to select redshifts around the median value of N2CO sources (z=2.8$\pm$0.1) and obtain ratios of  $S_{1.2\,mm}/S_{850\,\mu m}$=0.42 and $S_{2\,mm}/S_{850\,\mu m}$=0.082. These ratios apply only to DSFGs (i.e., not to radio sources). 
We are using them to compute the expected flux density at 850\,$\rm \mu$m (S$^{850}$) from our 1.2 and 2\,mm flux measurements. A source will be in the SCUBA2 catalog if it has S$^{850} \ge 4\sigma^{850}_{inst}$. We thus measured $\sigma^{850}_{inst}$ in the SCUBA2 map at each source position. Figure~\ref{fig:Expect_S850} shows a comparison between the 850\,$\rm \mu$m flux densities (for our $>$95\% purity sample) with the 4$\sigma^{850}_{inst}$ cut applied for source detection. Most of the sources lie close to the detection threshold, where the S2COSMOS completeness is $\sim$50\%. However, a few 2\,mm sources are clearly above the threshold (three of which are strong radio sources while one is a blended 1.2\,mm source) 
We repeated the procedure with the 80-95\% purity sample (Fig.~\ref{fig:Expect_S850}). In this case, N2CLS sources predominantly lie below the S2COSMOS detection threshold. For comparison, we also show the 4$\sigma^{850}_{\rm inst}$ corresponding to the deepest S2COSMOS region ($\sigma^{850}_{\rm inst}$=0.7\,mJy/beam). At such a depth all N2CLS sources should have been detected.

In summary, the absence of SCUBA2 counterparts for some N2CLS sources is primarily driven by sensitivity limits rather than intrinsic SED differences. Most of these sources lie close to the S2COSMOS detection threshold, and the fraction of non-detections increases toward lower N2CLS S/N, consistent with an effect driven by the 850\,$\mu$m detection limit. We find no evidence that SCUBA2 dropouts preferentially occur at higher redshift. 

So we can conclude that the current data do not require a distinct high-redshift or unusual-SED population to explain the SCUBA2 non-detections.
Rather, our millimeter survey appears to probe sources lying just below the effective completeness limit of S2COSMOS, and the discrepancy is expected to largely disappear at the depth of the deepest 850\,$\mu$m regions.

\section{Conclusion \label{sec:conclusion}}
We have presented the final NIKA2 observations of the COSMOS field obtained as part of N2CLS together with the associated 1.2 and 2\,mm maps and source catalogs. 
The survey provides deep and homogeneous millimeter coverage in the COSMOS field, enabling us to construct a robust sample of millimeter-selected DSFGs. The main results of our work are summarized as follows:
\begin{itemize}
\item We obtained the final N2CLS maps at 1.2 and 2\,mm covering $\sim$\,1070\,arcmin$^2$ in the COSMOS field with median noise levels of 
315\,$\mathrm \mu$Jy at 1.2\,mm and 91$\mathrm \mu$Jy at 2\,mm.

\item We constructed a master catalog of 323 millimeter-selected sources detected with S/N$\geq$3.9 (corresponding to $\geq$80$\%$ purity), including 104 sources detected in both bands, 197 detected only at 1.2\,mm, and 22 detected only at 2\,mm.

\item Multiwavelength counterpart identification using interferometric and radio data provided proxies for $\sim$89$\%$ of the sample, while only $\sim$11$\%$ of the sources remain without a secure counterpart.

\item The redshift distribution of the high-purity N2CLS galaxy sample at 1.2\,mm has a median redshift of z$\simeq$2.8, confirming that millimeter selection efficiently probes DSFGs near and beyond the peak epoch of cosmic star formation.

\item We compared the observed redshift distribution with predictions from several galaxy evolution models. While the SIDES simulations provide an excellent match to the N2CLS measurements, none of the other models fully reproduces the observed cumulative redshift distribution. In particular, although SHARK and the model of \cite{Zavala2021} predict median redshifts consistent with the observed value, they exhibit significant deviations in the shape of the distribution. The semi-empirical model of \cite{Popping+20}, which predicts a lower median redshift, shows an even larger discrepancy.

\item About 25\% of the N2CLS sources do not have a corresponding SCUBA2 850\,$\rm \mu$m detection, most likely due to the highly nonuniform noise distribution in the SCUBA2 map. This result further emphasizes the strength of the N2CLS survey, which uniquely combines wide-area coverage (0.3\,deg$^2$) with a highly uniform noise distribution across the entire field.

\end{itemize}

Beyond the specific results presented here, the primary goal of this work was to provide a high-quality legacy dataset for studies of dust-obscured galaxy evolution. The combination of the N2CLS millimeter observations with the extensive multiwavelength coverage of the COSMOS field makes this dataset particularly well suited for future studies of dust-obscured star formation across cosmic time. The released data products (Appendix\,\ref{app:data_release}) include the maps, blindly detected catalogs, simulations, and multiwavelength identifications, providing a well-characterized dataset for the community. These data will enable future investigations of the physical properties, environments, and cosmic evolution of DSFGs and will provide an important bridge between wide-area single-dish surveys and high-resolution interferometric studies.

\section*{Data availability}
The full catalog presented in Table\,\ref{tab: N2CLS_master} is publicly available at the CDS. The source cutouts are available at \url{https://doi.org/10.5281/zenodo.23361954}. The COSMOS field maps and NIKA2 catalogs are available through the IRAM Science Data Archive and the N2CLS data release page (\url{https://data.lam.fr/n2cls}), together with simulations of the COSMOS field, multiwavelength counterparts and redshift information for the N2CO sources. The N2CLS data release page is described in Appendix~\ref{app:data_release}.

\begin{acknowledgements}
We acknowledge financial support from the "Action Thématique de Cosmologie et Galaxies" (ATCG) funded by CNRS/INSU-IN2P3-INP, CEA and CNES, France, from the European Research Council (ERC) under the European Union's Horizon 2020 research and innovation program (project CONCERTO, grant agreement No 788212) and from the Excellence Initiative of Aix-Marseille University-A*Midex, a French ``Investissements d'Avenir'' program.

This work is based on observations carried out under project numbers 192-16 with the IRAM 30-m telescope, and project W23CJ with NOEMA. IRAM is supported by INSU/CNRS (France), MPG (Germany) and IGN (Spain).

We would like to thank the IRAM staff for their support during the NIKA and NIKA2 campaigns. The NIKA2 dilution cryostat has been designed and built at the Institut N\'eel. In particular, we acknowledge the crucial contribution of the Cryogenics Group, and in particular Gregory Garde, Henri Rodenas, Jean Paul Leggeri, Philippe Camus. This work has been partially funded by the Foundation Nanoscience Grenoble and the LabEx FOCUS ANR-11-LABX-0013 and has been supported by the French National Research Agency under the contracts "MKIDS," "NIKA," and ANR-15-CE31-0017 and in the framework of the "Investissements d’avenir” program (ANR-15-IDEX-02). This work has benefited from the support of the European Research Council Advanced Grant ORISTARS under the European Union's Seventh Framework program (Grant Agreement no. 291294). F.R. acknowledges financial supports provided by NASA through SAO Award Number SV2-82023 issued by the Chandra X-Ray Observatory Center, which is operated by the Smithsonian Astrophysical Observatory for and on behalf of NASA under contract NAS8-03060. 

MA is supported by FONDECYT grant number 1252054, and gratefully acknowledges support from ANID Basal Project FB210003,  ANID MILENIO NCN2024 112 and ANID + Vinculaci\'on Internacional + FOVI250261.

The NIKA2 data were processed using the Pointing and Imaging In Continuum (PIIC) software, developed by Robert Zylka and Stefano Berta at the Institut de Radioastronomie Millim\'etrique (IRAM) and distributed by IRAM via the GILDAS pages. PIIC is the extension of the MOPSIC data reduction software to the case of NIKA2 data.

CRCB acknowledges financial support from the French government under the France 2030 investment plan, as part of the Initiative d’Excellence d’Aix-Marseille Université – A*MIDEX AMX-22-RE-AB-101.

This work makes use of data from the S2COSMOS project (ID: M16AL002). We thank the S2COSMOS team for making these observations available. 

This work makes use of data from the A3COSMOS project. We thank the A3COSMOS team for their effort in collecting and reducing the ALMA observations, in particular Daizhong Liu, Benjamin Magnelli, and Silvia Adscheid for useful discussions about the catalogs.

We also warmly thank Olivier Ilbert for enlightening discussions about the use of COSMOS2020 and COSMOS2025 data.

\end{acknowledgements}

\bibliographystyle{aa} 
\bibliography{bibtex_cosmos} 

\begin{appendix}
\nolinenumbers 

\section{N2CLS noise maps}
\begin{figure}[!h]
\centering
\includegraphics[width=1\linewidth]{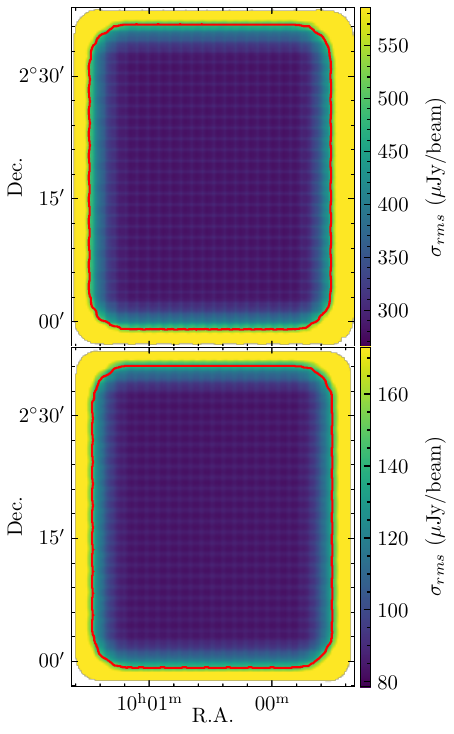}
\caption{N2CLS COSMOS 1$\sigma$ noise maps showing the point-source sensitivity at 1.2 (\textit{top}) and 2\,mm (\textit{bottom}). Units are $\mathrm{\mu Jy/beam}$.}
\label{fig:noise_maps}
\end{figure}

Figure~\ref{fig:noise_maps} shows the $1\sigma$ instrumental noise maps at 1.2 and 2\,mm. The red contour delineates the source extraction region, defined by the condition $\sigma_{\rm inst}<1.8\,\sigma_{\rm min}$, where $\sigma_{\rm min}$ is the minimum instrumental noise at the field center. The maps illustrate the noise homogeneity within this region.

\section{Building the master 1.2 and 2\,mm N2CLS source catalog \label{app:master_sep}}
We built an N2CLS master catalog  by cross-matching the blind catalogs extracted at 1.2\,mm and 2\,mm. We identified nine N2CLS sources for which the separation $r$ between the positions of the 1.2 and 2\,mm sources is larger than $6.5\arcsec$. These sources satisfy the matching criterion requiring the separation between the 1.2 and 2\,mm positions to be smaller than $r_{\rm matching}$. However, this does not guarantee that the 1.2 and 2\,mm sources correspond to the same galaxy. Therefore, we treated these cases carefully during the proxy identification in Sect.~\ref{sec: multiwave_ident}.

One clear example is N2CO\_1\_294, which has a separation distance of $r>9\arcsec$ and two distinct proxies for identification: one from A3COSMOS \citep{A3COSMOS}, matching the 1.2\,mm source, and another from the VLA 3\,GHz survey \citep{3GHZVLA}, matching the 2\,mm source. Therefore, we split this source into two sources, N2CO\_1\_294 ($\rm S/N_{1.2\,mm} = 3.9$) and N2CO\_2\_97 ($\rm S/N_{2\,mm} = 4.4$). N2CO\_2\_97 has a counterpart at 1.2\,mm with S/N$\sim$3.6, therefore below the 80\% purity threshold. We extracted its 1.2\,mm flux at the 2\,mm source position using prior-based photometry and corrected it with the transfer function.

We also separated N2CO\_1\_235 ($\rm S/N_{1.2\,mm} = 4.49$) from its 2\,mm counterpart N2CO\_2\_61 ($\rm S/N_{2\,mm} = 5.5$), which lies $\sim$8.8\arcsec\ away and is matched with a different proxy, offset from the 1.2\,mm source. A similar situation could occur for N2CO\_1\_81 and N2CO\_1\_284, where the separation is $r>$10\arcsec\ (close to the $r_{\rm matching}$ limit); however, we did not identify a distinct proxy for the 2\,mm counterparts, and therefore we did not separate them. Additionally, we identified two pairs of 1.2\,mm sources, N2CO\_1\_8-N2CO\_1\_15 and N2CO\_1\_41-N2CO\_1\_43, which share the same blended 2\,mm counterparts, $\rm ID_{2mm} = 6$ and 12, respectively.

Finally, N2CO\_2\_29 is blended, with a central (proxy) position $\sim$2.1\arcsec\ ($\sim$6.6\arcsec) away from the 1.2\,mm N2CO\_1\_153 source (proxy) position. The brightest pixels in the 2\,mm map correspond to a MeerKAT-identified source, with no 1.2\,mm counterpart. The 1.2\,mm source is identified with CHAMPS. Consequently, we consider them as distinct sources in the master catalog. 

\section{Photometric redshift comparison: COSMOS2020 versus COSMOSWeb}\label{app: photo_redshift}

In this appendix we assess the performance of COSMOS2020 and COSMOSWeb photometric redshift catalogs for the N2CLS galaxy sample by comparing photometric and spectroscopic redshifts. To quantify the accuracy of the photometric redshifts, we used the normalized redshift difference:
\begin{equation}
\Delta z = \frac{z_{\mathrm{phot}} - z_{\mathrm{spec}}}{1 + z_{\mathrm{spec}}}.
\end{equation}
The median of this quantity, $\langle \Delta z \rangle$, defines the photometric redshift bias and measures the presence of a systematic offset between photometric and spectroscopic redshifts. The scatter is characterized using the normalized median absolute deviation, defined as:
\begin{equation}
\sigma_{\mathrm{NMAD}} = 1.48 \times \mathrm{median}\left( \left| \Delta z - \langle \Delta z \rangle \right| \right),
\end{equation}
which provides a robust estimate of the dispersion that is insensitive to outliers. Catastrophic outliers are defined as sources with $|\Delta z| > 0.15$.

\begin{figure}
\centering
\includegraphics[width=0.9\columnwidth]{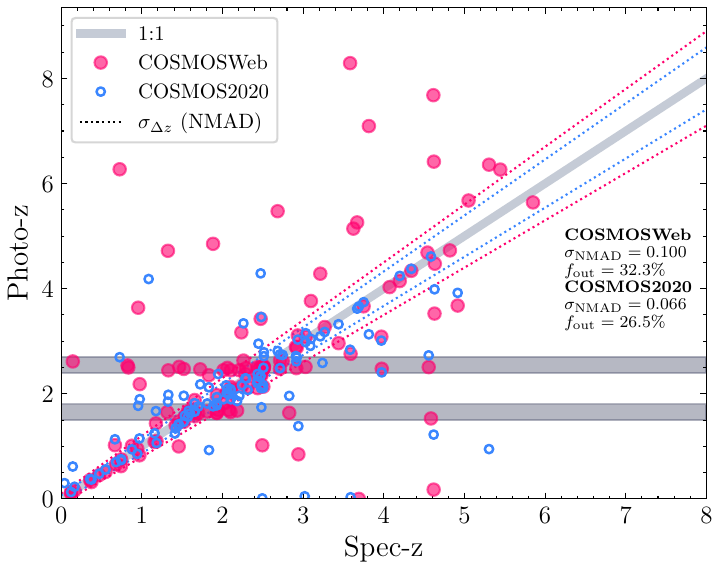}
\caption{Comparison between spectroscopic and photometric redshifts for the N2CLS galaxies using the COSMOSWeb and COSMOS2020 catalogs. The solid gray line indicates the one-to-one relation, while the dotted lines represent the $\sigma_{\Delta z}$ envelopes derived from the normalized median absolute deviation (NMAD). COSMOSWeb photometric redshifts show larger scatter, a higher fraction of catastrophic outliers, and artificial overdensities at z=1.6 and z=2.5 (gray shaded regions) compared to COSMOS2020.}
\label{fig: redshift_photo_vs}
\end{figure}

The comparison between photometric and spectroscopic redshifts reveals systematic differences between the COSMOS2020 and COSMOSWeb catalogs (see Fig.~\ref{fig: redshift_photo_vs}). Both catalogs show the same median bias, $\langle \Delta z \rangle = -0.004$, indicating the absence of a significant global offset between $z_{\mathrm{phot}}$ and $z_{\mathrm{spec}}$. However, the COSMOSWeb photometric redshifts exhibit a larger scatter, $\sigma_{\mathrm{NMAD}} =0.100$ compared to $\sigma_{\mathrm{NMAD}} = 0.066$ for COSMOS2020, as well as a higher fraction of catastrophic outliers, $f_{\mathrm{out}} = 32.3\%$ versus $26.5\%$, respectively. These results indicate that, despite comparable average bias, the COSMOSWeb catalog is more affected by source-to-source uncertainties for the N2CLS galaxy population.

Furthermore, the redshift distribution inferred from COSMOSWeb photometric redshifts displays pronounced over-densities in two redshift bins, centered at z=1.6 and z=2.5, with excess fractions of 35\% and 40\% relative to the spectroscopic distribution. This excess is not driven by the underlying spectroscopic redshift distribution, but instead reflects systematic effects in the photometric redshift estimation. The presence of these artificial peaks, together with the increased $\sigma_{\mathrm{NMAD}}$ and outlier fraction, provides strong evidence of residual biases in the COSMOSWeb photometric redshifts when applied to dusty millimeter-selected galaxies. For this reason, COSMOS2020 photometric redshifts are adopted as the reference throughout this work whenever available.

\section{N2CLS and SIDES redshift distributions for 2\,mm sources \label{App:2mm_z}}
We extended the comparison between N2CLS and SIDES redshift distributions to the 2\,mm sources at 95\% purity (S/N$\ge$4.6). Compared to \citetalias{Bethermin+26}, we collected high-resolution (sub)millimeter observations from more catalogs (CHAMPS and NOEMA) and this improves the redshift completeness from 88\% to 97\%. At this purity level, the sample includes 90 2\,mm sources, which break into 92 N2CLS sources: 69 single-component systems, 18 multi-component systems resolved into 43 counterparts, and 3 N2CLS sources without identification. As discussed in Appendix~\ref{app:master_sep}, two pairs of sources share the same 2\,mm counterpart, namely N2CO\_1\_8–N2CO\_1\_15 (ID$_{\rm 2\,mm}$=6) and N2CO\_1\_41–N2CO\_1\_43 (ID$_{\rm 2\,mm}$=12). For these pairs, we assign the redshift of the brightest component, following the same procedure adopted for multi-component sources. The final subsample comprises 87 sources with assigned redshifts (69 single-component and 18 multi-component systems), with the remaining 3 sources lacking redshift information due to the absence of an identified high-resolution counterpart. To enable a direct comparison with the SIDES catalogs, we applied the same flux limit as for the 1.2\,mm subsample, adopting $\rm S_{2\,mm} \geq S/N \times \sigma_{\rm noise~2\,mm} = 0.42$\,mJy. 
The median redshift of the 2\,mm high-purity sample is z=3.0$\pm$0.2. This is consistent with \citetalias{Bethermin+26}, who report a mean redshift of z=3.2$\pm$0.2, in agreement with the mean of our updated sample.

SIDES predicts median redshifts of 3.0$\pm$0.2 for the galaxy catalog and 3.0$\pm$0.1 for the blob catalog, in perfect agreement with the N2CLS measurements.  
The KS test gives $p$-values of 0.23 and 0.43 for the blob and galaxy catalogs, respectively, with a maximum deviation of $D \sim 0.1$ at z$\sim$4, where the cumulative N2CLS distribution lies slightly below that of SIDES, with a difference consistent with statistical fluctuations. 
These results support the hypothesis that the samples are drawn from the same parent distribution, as also found for the 1.2\,mm high-purity sample.

As for the 1.2\,mm sample, we verified that these results are not driven by 
the redshift uncertainties, repeating the KS test over $10^3$ Monte Carlo 
realizations of the 2\,mm redshift distribution following the scheme 
described in Sect.~\ref{sec: z_distribution_sides}. We obtain 
$D = 0.098^{+0.017}_{-0.017}$ and $p = 0.35^{+0.24}_{-0.17}$ for the blob catalog, and $D = 0.088^{+0.016}_{-0.013}$ and $p = 0.49^{+0.20}_{-0.20}$ 
for the galaxy catalog, consistent with the values obtained from the 
unperturbed redshifts. The null hypothesis is rejected ($p$-value$<0.05$) in 
less than $1\%$ of the realizations for both catalogs.

\section{Comparison of the N2CLS and SCUBA2 850\,$\rm \mu$m surveys}\label{app: scuba2}

\begin{figure}[h]
    \centering
     \includegraphics[width=0.46\textwidth]{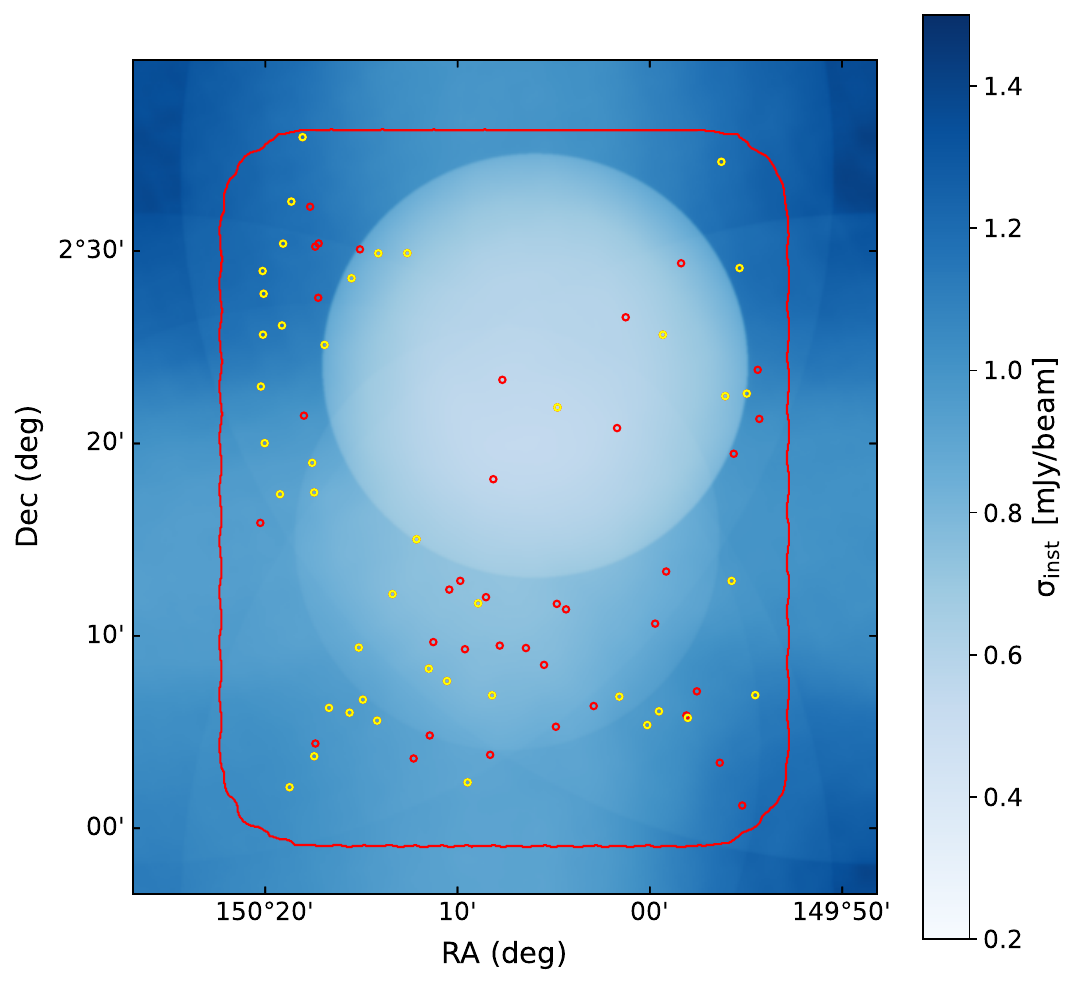}
    \caption{Location of the N2CLS sources without SCUBA2 counterparts (yellow: $\ge95$\% purity sample; red: purity between 80\% and 95\%). The map shows the match filtered instrumental noise map at 850\,$\mathrm \mu$m from S2COSMOS \citep{Simpson+19}. The red contour corresponds to the area where the N2CLS sources are extracted and is the same as in Fig.\,\ref{fig: maps_1.2_2_mm}.}
    \label{fig:N2CLS_no850}
\end{figure}

Additional information on N2CLS sources without SCUBA2 counterparts is provided here. The locations of these sources are shown in Fig.~\ref{fig:N2CLS_no850} on the S2COSMOS 850\,$\rm \mu$m instrumental noise map. Tables~\ref{tab: numbers_high} and \ref{tab: numbers_low} list the corresponding numbers of sources in the different subsamples. Most of these sources lie outside the very deep central region of the SCUBA2 coverage.

\begin{table}[!h]
\centering
\caption{Number of N2CLS sources without a SCUBA2 counterpart for the $\geq$95\% purity sample (S/N$\geq$4.5 at 1.2\,mm, S/N$\geq$4.6 at 2\,mm).}
\label{tab: numbers_high}
\begin{tabular}{cccc}
\hline
 & N2CLS & No S2COSMOS & No SCUBA2\\\hline
1.2 \& 2\,mm  & 101   & 3  & 2 \\
1.2\,mm           & 132  &  35 (1) & 33 (1) \\
2\,mm             & 10   &  7 & 7 \\ \hline
TOTAL & 243   &  45  (1) & 42  (1) \\ \hline
\end{tabular}
\tablefoot{S2COSMOS refers to the SCUBA2 survey presented in \cite{Simpson+19}; SCUBA2 includes in addition the STUDIES survey. 
In parenthesis are the number of N2CLS sources with multiple identifications, all the other sources have single counterpart or no identification.}
\end{table}

\begin{table}[!h]
\centering
\caption{Same as Table~\ref{tab: numbers_high} for N2CLS sources with purity between 80\% and 95\% (3.9$<$S/N$<$4.5 at 1.2\,mm and 3.9$<$S/N$<$4.6 at 2\,mm).}
\label{tab: numbers_low}
\begin{tabular}{cccc}
\hline
& N2CLS & No S2COSMOS & No SCUBA2 \\\hline
1.2 \& 2\,mm   &  3 & 0 & 0 \\
1.2\,mm           & 65  & 35 (2) & 30 (2)\\
2\,mm             & 12 & 8 & 7 \\ \hline
TOTAL & 80 & 43 (2) & 37 (2)\\ \hline
\end{tabular}
\end{table}

\section{Data release \label{app:data_release}}
We are releasing on \url{https://data.lam.fr/n2cls/data} the following datasets:
\begin{itemize}
\item The N2CLS signal maps (in mJy/beam) and weight (inverse variance of the instrument noise) maps (Sect.~\ref{subsec: data_redu}).
\item The blind catalogs at 1.2\,mm (ar13) and 2\,mm (ar2) (Sect.~\ref{sec: N2CLS_catalog}).
\item The complete set of 117 simulations, including the input galaxy catalogs from SIDES, the input maps (that are then used to construct NIKA2 timelines), and the output maps after passing through the entire data reduction pipeline (Sect.~\ref{sect:cat_char}).
\item The N2CLS merged 1.2 and 2\,mm S/N$\ge$3.9 N2CO master catalog (Sect.~\ref{sec: N2CLS_catalog}).
\item The identifications and redshifts of all sources of the N2CO catalog (Sect.~\ref{sec: proxy_identification} and \ref{sec:n2cls_galaxies}).
\item The multiwavelength counterparts of all sources of the N2CO catalog (Sect.~\ref{sect: SED}).
\end{itemize}

\begin{figure}[!t]
\centering
\includegraphics[width=0.85\linewidth]{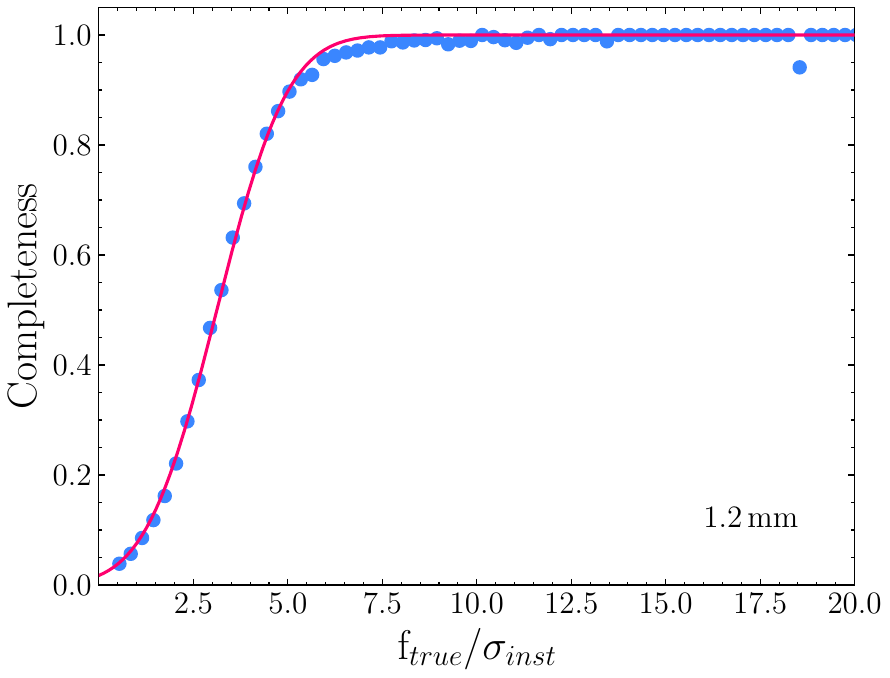}
\includegraphics[width=0.85\linewidth]{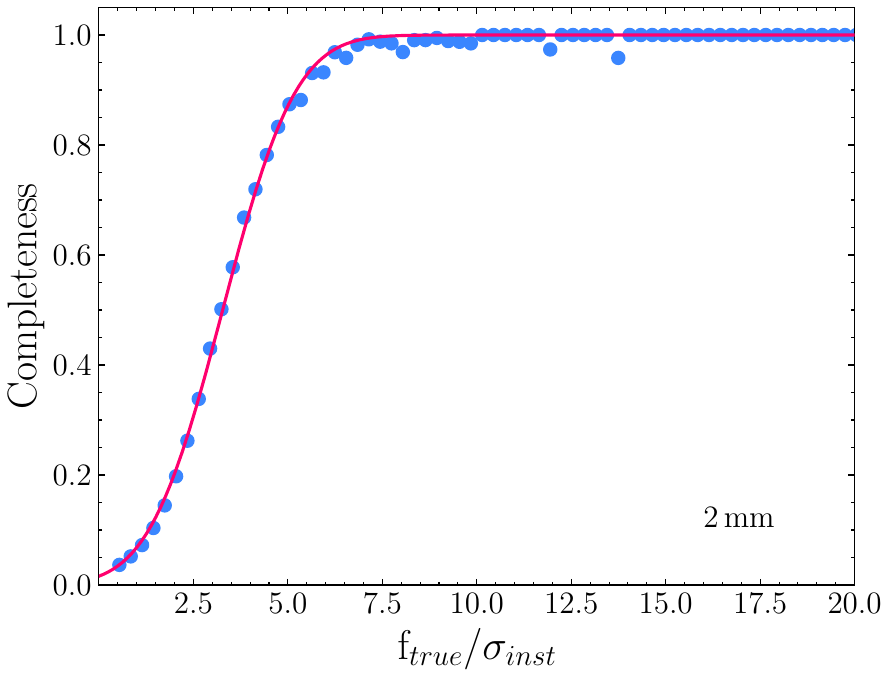}
\caption{Source completeness as a function of the ratio between the input flux from the blob catalog and instrumental noise level ($f_{true}/\sigma_{inst}$) at 1.2\,mm (\textit{top}) and 2\,mm (\textit{bottom}). The red line shows the best fit using the Eq.~\ref{eq: completeness}.}
\label{fig: completeness}
\end{figure}

The 1.2 and 2\,mm catalogs also include information on completeness and effective area, which will be useful for statistical analyses.
Completeness quantifies the probability that a source from the noiseless sky maps (the ``blob catalog'') is recovered in the output catalog, as a function of its intrinsic properties.
As an intrinsic property, we only consider the input flux density of the blob sources ($f_{true}$). To account for the small instrumental noise inhomogeneities in the survey, we also consider the instrument noise level at the position of the source ($\sigma_{inst}$). Therefore, the completeness is evaluated as the fraction of input blob sources at a given $f_{true}/\sigma_{inst}$ which are recovered in the output ``simulated catalog.'' We fit the results using
\begin{equation}
C(x)=\frac{1+\operatorname{erf}\left(\frac{f_{true}/\sigma_{inst}-x_0}{y_0}\right)}{2},
\label{eq: completeness}
\end{equation}
where $C$ is the completeness, $x_0$ and $y_0$ are free parameters. In Fig.~\ref{fig: completeness}, we present the results of our measurements and the fitted curve. 

We also derived the effective area of the survey at a given flux density, to take into account the fact that faint sources are unlikely to be detected in noisy regions of the field. The effective area is defined as the sum of the surface area of each pixel ($\Omega_{pix}$) multiplied by the completeness at the pixel position at a given flux density \citep[see][]{Bethermin+20, Bing+23}:
\begin{equation}
\Omega_{\mathrm{eff}}(S)=\sum_{i=1}^{N_{\text{pixel}}} \Omega_{\mathrm{pix}} \times \mathcal{C}\left(\frac{S}{\sigma_i}\right),
\label{eq: eff_area}
\end{equation}
where $\sigma_{i}$ is the instrumental noise at the $i$th pixel, C is the completeness function (Eq.~\ref{eq: completeness}) and $S$ the true flux density of a given source (i.e., the corrected flux, $\rm S_{corr}$, defined in Sect.~\ref{sec: N2CLS_catalog}, and corrected using the effective flux boosting factor derived in Sect.~\ref{sec: flux_correction}). We thus determined $\Omega_{\rm eff}$ for each detected source taking into account the nonuniform noise level of the maps. Sources with low flux densities have smaller effective area.

\section{Example cutouts of detected sources \label{app:cutouts}}
\begin{figure*}[!t]
\centering
\newlength{\figwidth}
\setlength{\figwidth}{0.8\textwidth}
\makebox[\figwidth][l]{\includegraphics[width=0.76\textwidth]{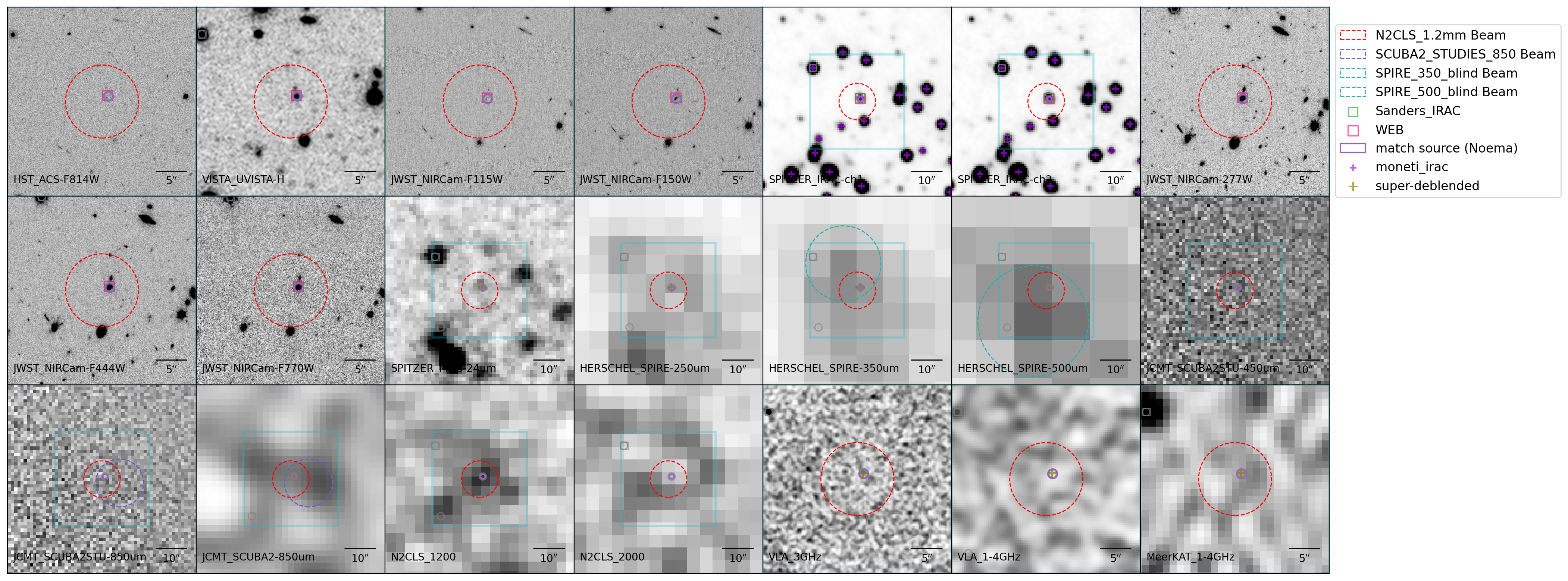}} \\[2pt]
\makebox[\figwidth][l]{\includegraphics[width=0.77\textwidth]{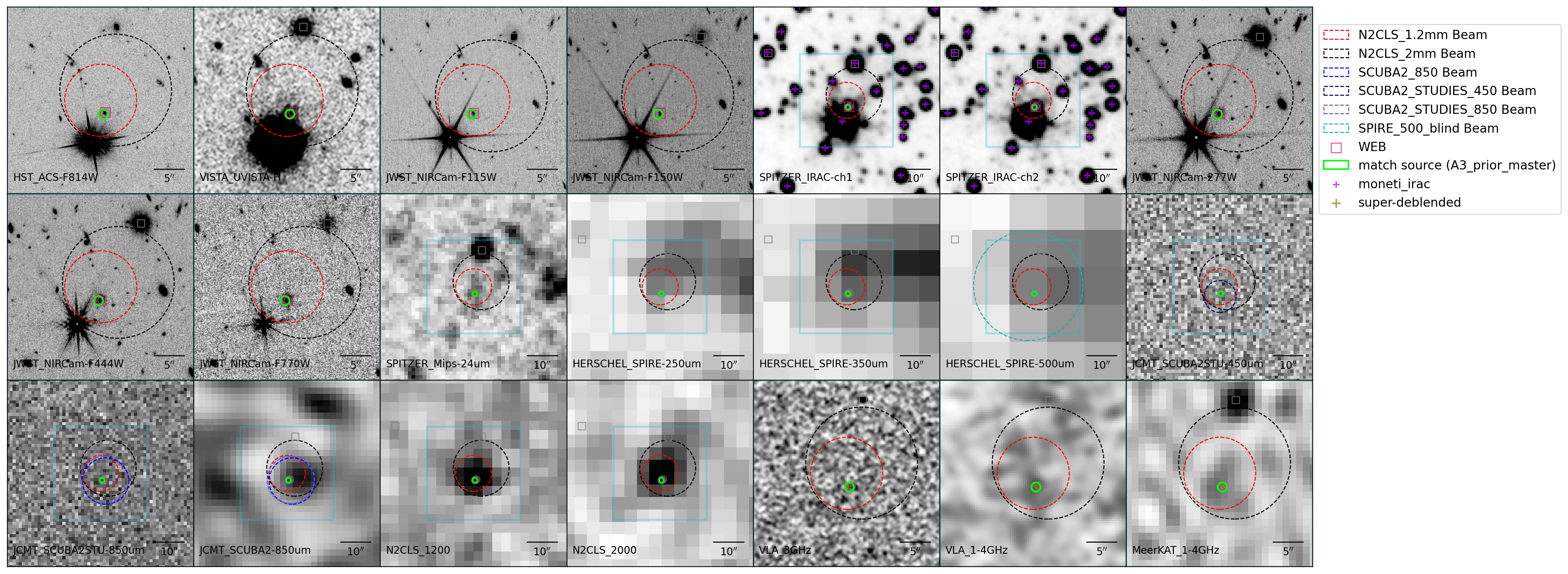}} \\[2pt]
\makebox[\figwidth][l]{\includegraphics[width=0.77\textwidth]{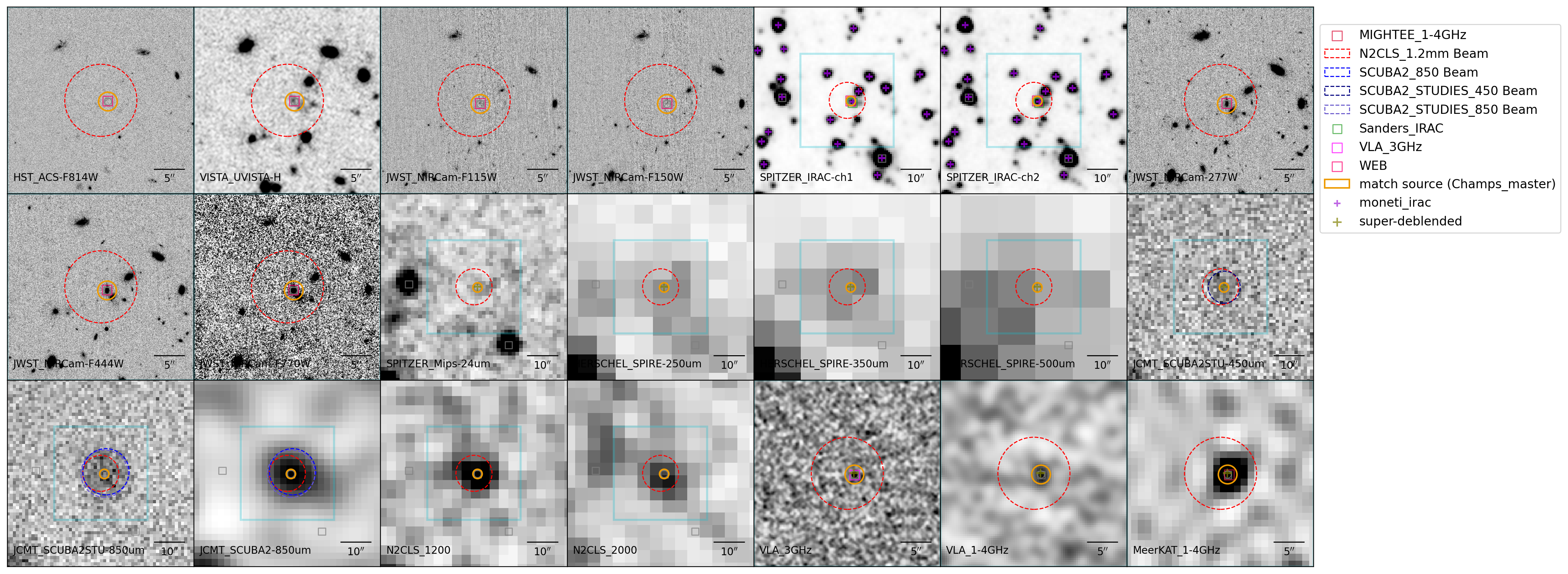}} \\[2pt]
\makebox[\figwidth][l]{\includegraphics[width=0.755\textwidth]{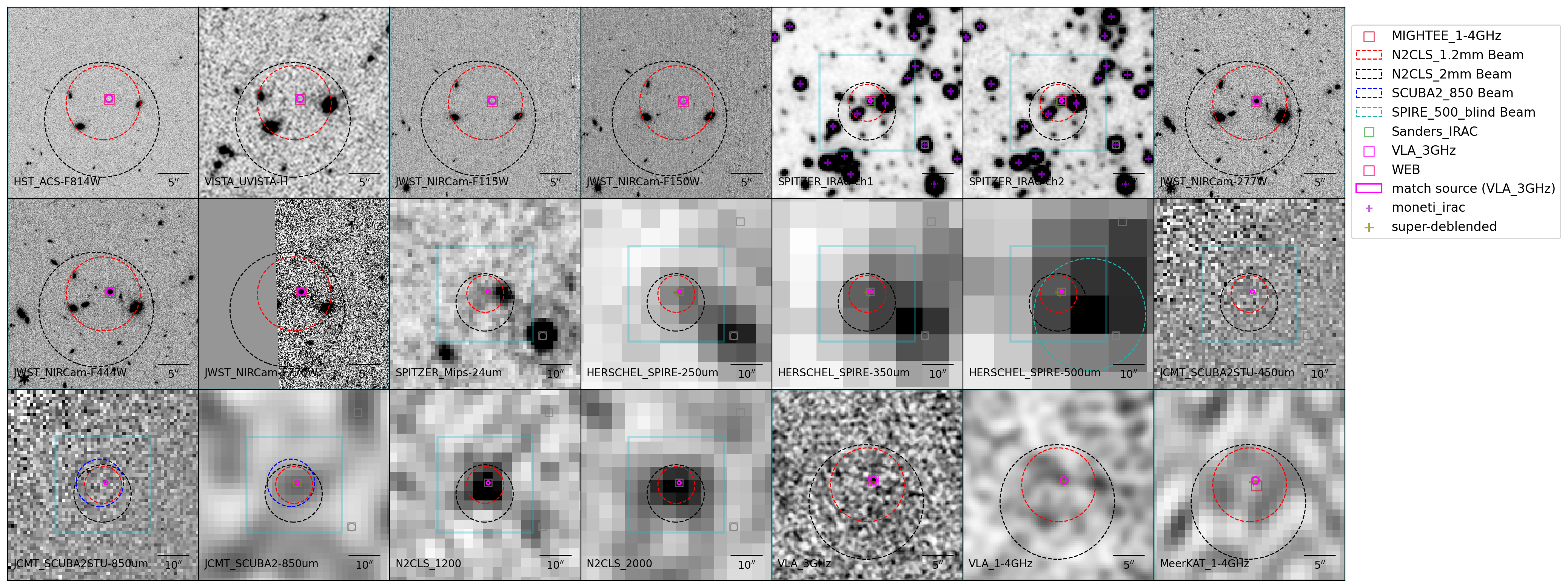}}
\caption{Multiwavelength cutouts of four N2CLS sources in the COSMOS field. The identified proxy (NOEMA, A3COSMOS, CHAMPS, or VLA\,3\,GHz) is shown as a colored circular beam with radius $\mathrm{FWHM}/2$. Other high-resolution observations are indicated by gray markers (circles, squares, or diamonds). The beam sizes of far-IR and (sub)mm observations (e.g., SPIRE, SCUBA2, and NIKA2) are represented by dashed circles with radius $\mathrm{FWHM}/2$. Optical, NIR, mid-IR, and radio catalogs are marked with squares. The super-deblended \citep{Jin+18,Liu+18} and IRAC \citep{Moneti+22} catalogs are indicated with plus symbols. The sizes of the cutouts are $15\times15$\,arcsec for optical and radio images, and $30\times30$\,arcsec from NIR to mm; the light cyan square highlights the optical cutout region. From top to bottom: N2CO\_1\_92 (NOEMA), N2CO\_1\_47 (A3COSMOS), N2CO\_1\_56 (CHAMPS), and N2CO\_1\_78 (VLA\,3\,GHz). The cutouts of the full sample are available at \url{https://doi.org/10.5281/zenodo.23361954} and \url{https://data.lam.fr/n2cls/data}.}
\label{fig: example_cutout}
\end{figure*}

Figure~\ref{fig: example_cutout} shows a subset of the available multiwavelength source cutouts for illustration. The full set comprises 371 cutouts: 337 associated with identified N2CO sources (including multiple counterparts) and 34 corresponding to unidentified N2CO sources. These cutouts, together with ALMA cutouts from A3COSMOS when available, can be accessed on the N2CLS data release page.

\end{appendix}

\end{document}